\documentclass[twocolumn,aps,prb,showpacs]{revtex4-1}
\usepackage{amsmath,amssymb}
\usepackage{graphicx}
\usepackage{psfrag}
\usepackage{epstopdf}
\usepackage{braket}
\usepackage[utf8]{inputenc}
\usepackage{mhchem}
\usepackage{natbib}

\usepackage{xspace}

\newcommand*{\eg}{\emph{e.g.}\@\xspace}
\newcommand*{\ie}{\emph{i.e.}\@\xspace}
\newcommand*{\cf}{\emph{cf.}\@\xspace}

\newcommand*{\etc}{%
    \@ifnextchar{.}%
        {etc}%
        {etc.\@\xspace}%
}

\begin{document}
\title{A Practical Guide to Surface Kinetic Monte Carlo Simulations}
\author{Mie Andersen}
\email{mie.andersen@ch.tum.de}
\affiliation{Chair for Theoretical Chemistry and Catalysis Research Center, Technische Universit{\"a}t M{\"u}nchen,\\ Lichtenbergstr.\ 4, D-85747 Garching, Germany}
\author{Chiara Panosetti}
\affiliation{Chair for Theoretical Chemistry and Catalysis Research Center, Technische Universit{\"a}t M{\"u}nchen,\\ Lichtenbergstr.\ 4, D-85747 Garching, Germany}
\author{Karsten Reuter}
\affiliation{Chair for Theoretical Chemistry and Catalysis Research Center, Technische Universit{\"a}t M{\"u}nchen,\\ Lichtenbergstr.\ 4, D-85747 Garching, Germany}

\date{\today}

\begin{abstract}
This review article is intended as a practical guide for newcomers to the field of kinetic Monte Carlo (KMC) simulations, and specifically to lattice KMC simulations as prevalently used for surface and interface applications. We will provide worked out examples using the {\tt kmos} code, where we highlight the central approximations made in implementing a KMC model as well as possible pitfalls. This includes the mapping of the problem onto a lattice and the derivation of rate constant expressions for various elementary processes. Example KMC models will be presented within the application areas surface diffusion, crystal growth and heterogeneous catalysis, covering both transient and steady-state kinetics as well as the preparation of various initial states of the system. We highlight the sensitivity of KMC models to the elementary processes included, as well as to possible errors in the rate constants. For catalysis models in particular, a recurrent challenge is the occurrence of processes at very different timescales, \eg fast diffusion processes and slow chemical reactions. We demonstrate how to overcome this timescale disparity problem using recently developed acceleration algorithms. Finally, we will discuss how to account for lateral interactions between the species adsorbed to the lattice, which can play an important role in all application areas covered here.
\end{abstract}

\maketitle

\section{Introduction}

As the witty name suggests, Monte Carlo is a wide umbrella term that covers a numerous family of approaches with one simple central idea in common: the resolution of complex problems using random numbers. Given the versatility of the concept, it is no surprise that Monte Carlo based approaches have gained popularity in computational chemistry and materials science (\cf \eg \cite{Frenkel}), most prominently for the simulation of ensemble properties using Metropolis Monte Carlo, or methods derived from the latter such as Basin Hopping for global geometry optimization. In addition to \emph{equilibrium} properties, the Monte Carlo idea can also be exploited to tackle \emph{dynamical} properties. In this sense, a number of approaches emerged over the decades under different names, until the term kinetic Monte Carlo (KMC) became universally used in this context.

Nowadays KMC is a popular tool to describe a variety of phenomena related to \eg transport (diffusion), structures and properties of materials (\eg crystal growth) or equilibrium and non-equilibrium chemistry (catalysis). As will become apparent throughout the text, in the context of atomistic simulations KMC can be considered as a form of coarse graining. This renders it particularly suitable to find its place in hierarchical multiscale modeling approches, where information at different levels of accuracy or detail is integrated to provide a more comprehensive description. In this context, KMC is an essential method to bridge the gap between the microscopic world (elementary processes such as atomistic diffusion jumps or the making and breaking of chemical bonds) and the meso- to macroscopic world (\eg a diffusion constant or a reaction rate).

\begin{figure*}
\centering
\includegraphics*[width=1.0\textwidth]{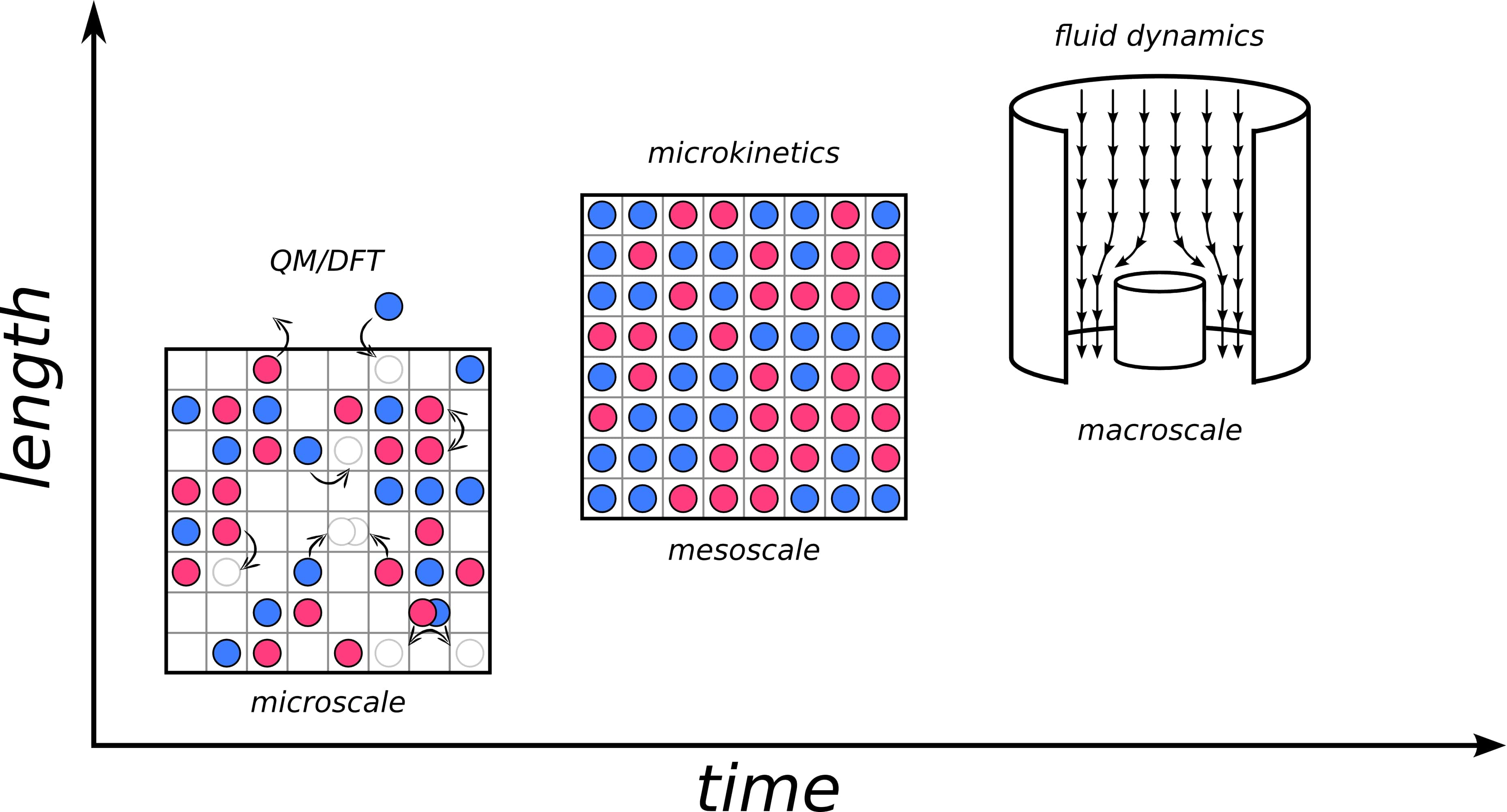}
\caption[]{Diagram of three scales involved in a hierarchical multiscale approach to heterogeneous catalysis and the corresponding models and methods. Bottom left: microscale ($\sim$nm in length, $\sim$ns in time). The schematic top view of a catalyst surface shows elementary processes like adsorption and desorption of chemical species, diffusion or reactive events in form of pictorial sketches involving the motion of spheres on a square lattice of active sites. Center: mesoscale ($\sim$$\mu$m in length, $\sim$ms in time). Microkinetic simulations evaluate the interplay of all the elementary processes on the microscale to yield information on surface composition (illustrated by the shown coverage of the catalyst surface with surface species) and intrinsic catalytic activity. Top right: macroscale ($\sim$m in length, $\sim$minutes/hours in time). Illustration of the fluid flow in a stagnation point reactor where the intrinsic catalytic activity determined by the microkinetic model enters as a boundary condition at all finite element cells describing the catalyst surface.}
\label{fig:multiscale}
\end{figure*}

Let us illustrate this concept by considering for example heterogeneous catalysis, which is one of the fields where KMC, as well as hierarchical approaches in general, have made considerable impact \cite{Sabbe2012,Reuter2016}. The design of a catalytic process requires a deep understanding of phenomena ranging from the reactive chemistry to the optimization of heat and mass transport within the reactor. Numerical simulations have become an integral part in this design process, which requires appropriate models at multiple time and length scales, and perhaps most importantly, concepts on how these models should be connected. One can imagine ``zooming in'' from the macroscopic world where we live and an industrial reactor operates, all the way down to the microscopic scale where the events are ultimately governed by electronic structure: adsorption and desorption of atoms and molecules at surfaces, diffusion, bond breaking and bond forming (see Fig.\ \ref{fig:multiscale}). All of the latter constitute elementary processes that can occur at the interface between the catalyst and the reaction fluid, and can nowadays be described individually to a great level of detail by first-principles electronic structure calculations. At this scale, what one essentially needs is a mechanistic description in terms of the Potential Energy Surface (PES) of the system ({\em vide infra}). Hereby, an appropriate quantum mechanical approach is required to capture chemical subtleties in a predictive manner. The workhorse for this remains to this day largely Density-Functional Theory (DFT) \cite{DFT1,DFT2}, thanks to its unique compromise between accuracy and efficiency that allows to access system sizes as relevant for heterogeneous catalysis.

If one now ``zooms out'' a little and into the mesoscopic scale, one can see how what globally happens is the result of an intricate interplay of the above elementary processes as well as the interaction with the surrounding environment. Here, the spatiotemporal evolution at the interface is dominated by collective behavior; dynamics and thermodynamics come into play. At this stage, one may employ the microscopic information (\eg reaction barriers, adsorption energies etc. of the elementary processes) and embed it into microkinetic models as a form of coarse-graining. A plethora of approaches of different complexity are available, from Sabatier analysis to mean-field models and kinetic Monte Carlo. Finally, at the macroscale, one needs to take into account transport phenomena and gradients of mass and temperature in the reactor geometry. This is a realm that is presently largely covered by continuum-type fluid dynamical models \cite{Janardhanan2011}. At this stage, again the information from the lower scales can be embedded, \ie here the outcome of the microkinetic models is integrated as an input, for instance in form of a boundary condition to the differential equations describing the flow phenomena \cite{Matera2010,Matera2014}. The development of appropriate hierarchical models to effectively describe events at such different scales, and more so of ``bridges'' to transfer information between them, constitutes the core of modern multiscale modeling approaches (\cf Ref.\ \cite{Raimondeau2002,Reuter2005,Vlachos2012}).

Similar kind of multiscale approaches have been or are being developed in many other areas of chemistry and materials sciences. Microkinetic models, and there prominently KMC simulations, are generally the approach used to capture the statistical interplay between elementary processes whenever the mesoscopic property or functionality to be described is outside of thermodynamic equilibrium. Besides catalysis, notable such application areas with a similar focus on surfaces or interfaces are diffusion and crystal growth. In this understanding of the use of KMC simulations we will concentrate in the following on this particular technique in these particular application areas. The present is, however, not intended as yet another extensive account of the fundamental methodology. For this we refer the reader to the many excellent reviews available in literature \cite{Voter,Chatterjee2007,Reuter2011,Stamatakis2012,Stamatakis2014}. Instead, what we want to provide is a practical guide of how to carry out such simulations (especially in the context of surface kMC), with particular emphasis on best practice recommendations as well as a discussion of current challenges and perspectives.

\section{KMC simulations: from theory to codes}

\subsection{Rare-event dynamics: a bottleneck which enables its own solution}
\label{sec:rare-events}

\begin{figure*}
\centering
\includegraphics[width=1.0\textwidth]{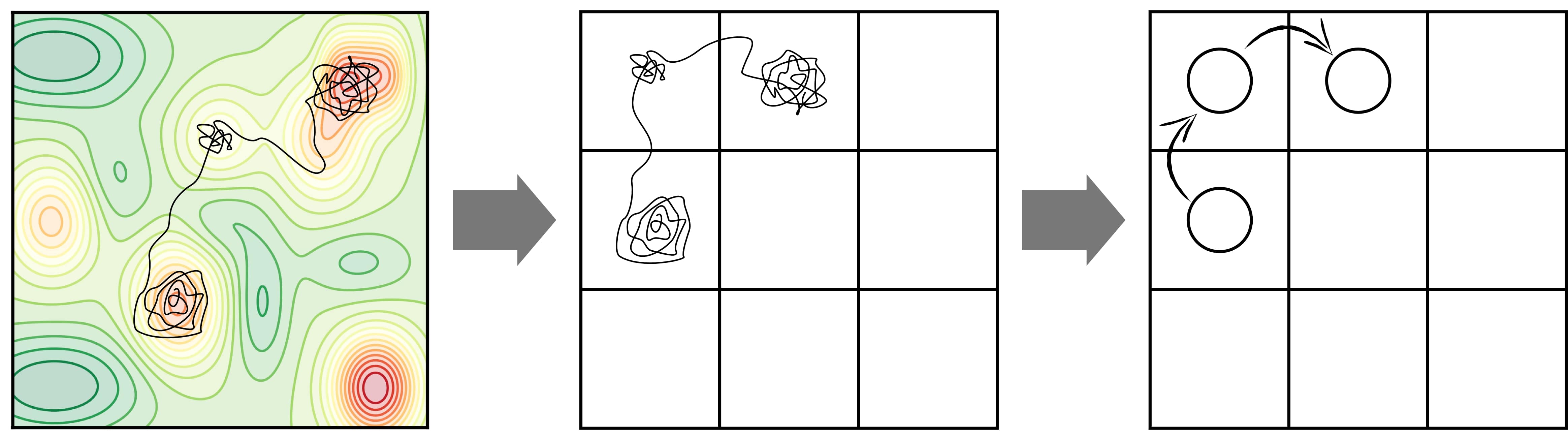}
\caption[]{Coarse-graining of a molecular dynamics (MD) trajectory into a Markov chain. Left: a possible MD trajectory (black) overlayed on the underlying potential energy surface (PES) of the system with red regions representing lower-energy basins. A large fraction of time is spent in these PES basins in vibrational motion around the respective minimum. At a certain moment in time, the systems finds an escape route to the next basin. Center: Coarse-graining of PES minima into positions on a suitably defined lattice. Each lattice position represents the basin of attraction of a PES miminum. Right: Coarse-graining of the continuous MD trajectory into a Markov chain of discrete hops between the basins/lattice positions.}
\label{fig:coarse_grain}
\end{figure*}

Many elementary processes involved at surfaces of solids exhibit high activation barriers (even despite a possible reduction of the barrier due to the presence of the catalyst in heterogeneous catalysis). These barriers are usually much larger than $k_BT$ and the corresponding processes are thus classified as rare events, if only thermal energy is there to drive them. While the motion of individual atoms (\eg their vibrations, but also the actual reaction events, \ie crossing an activation barrier once the system has reached the transition state) occurs on picosecond time scales, the time between consecutive high-barrier events can therefore be many orders of magnitude longer, possibly requiring simulations up to seconds or more in order to arrive at meaningful conclusions about the effect of the statistical interplay within an ensemble of multiple possible elementary processes. The ``life'' of our system in the long time span between these rare events is filled with vibrational motion around a single minimum on the PES. The relevant transitions to other (meta)stable states aka PES basins occur only occasionally. On a mesoscopic time scale, the time evolution of our system therefore manifests itself as a series of consecutive jumps from state to state (see Fig.\ \ref{fig:coarse_grain}). Additionally, it is intuitive to assume that the longer the time the system spends in one basin, the more it ``forgets'' how it actually got there. After a while, each possible way to escape from the basin therefore becomes completely independent of the entire preceding history before entering the basin. In other words, the state-to-state jumps of the system constitute a so-called Markov chain (\cf \eg \cite{vankampen2011}).

In consequence, the change of the probability $P_i(t)$ of the system to actually be in state $i$ at time $t$ depends only on the probabilities of hopping out of the current state $i$ into any other state $j$, $k_{ij}$, and on the probabilities of hopping into state $i$ from any other state $j$, $k_{ji}$. In the present context of chemical kinetics, these hopping probabilities are expressed as rate constants of the elementary processes with units time$^{-1}$. The overall change in $P_i(t)$ is thus governed by a simple balancing equation, called a Markovian master equation, that only contains these rate constants:
\begin{equation}
\frac{\mathrm{d}P_i(t)}{dt} = -\sum_{j\neq i} k_{ij} P_i(t) + \sum_{j\neq i} k_{ji} P_j(t) \quad .
\label{eqn:master}
\end{equation}
From a mathematical standpoint, Eq.\ \ref{eqn:master} is a system of coupled differential rate equations. Seemingly simple, it unfortunately becomes quickly unfeasible to solve explicitly for the number of possible states typically involved in surface catalysis (or diffusion or growth). For a rough estimate, let us consider a system with 100 surface sites (\eg an fcc(100) slab of ten atoms per edge in an otherwise periodic boundary condition cell to simulate an extended surface). In the course of a KMC simulation modeling a simple catalytic reaction A + B $\rightarrow$ AB, each of these sites can assume 3 possible occupation states; it can be empty, occupied by species A or occupied by species B (assuming that a formed product AB immediately desorbs into the gas phase). Already the number of possible configurations of such a trivial toy system is then $3^{100}$! The matrix $k_{ij}$ containing all the possible rate constants between system states will thus have $(3^{100})^2 \approx 2.66\ 10^{95}$ elements, making even its comprehensive storage impossible, let alone its diagonalization -- and even considering that the limited number of accessible states renders it largely sparse.

As an ingenious way out of this mess, the practical Monte Carlo type idea behind KMC is to never even attempt to deal with the entire matrix, but instead to generate stochastic trajectories that propagate the system from state to state (\ie a Markovian sequence of discrete hops to random states happening at random times). From this, the correct time evolution of the probabilities $P_i(t)$ is then obtained by ensemble averaging over these trajectories, or, if the system is in a steady state and ergodicity is ensured, by time averaging over a singular, sufficiently long trajectory. The KMC method thus replaces the analytical solution of Eq.\ \ref{eqn:master} with a numerical approach based on stochastic dynamics. The "only" thing that is needed to make this work is an algorithm that generates suitable such trajectories that (once ensemble or time averaged) yield the correct probabilities $P_i(t)$. This algorithm hence needs to determine at each step along such a state-to-state trajectory to which state the system should jump next and after what time step this next jump should happen. The KMC algorithm does so by selecting elementary processes according to their rate constants, followed by an updating of the time. We will come back to these algorithmic details in Sec.\ \ref{algo}.\\

\subsection{Mapping the problem onto a lattice}
\label{sec:lattice}

The challenges in applying KMC in practice are largely connected to the plethora of minima on the relevant PESs, and more so, the even larger number of elementary processes connecting them \cite{Margraf2019}. In the initially sketched multiscale view it would be desirable to employ rate constants calculated from first principles for all of these processes in order to establish models of predictive power. If one considers that each rate constant calculation requires in principle the determination of a transition state to get the activation barrier ({\em vide infra}), a brute-force KMC approach that requires at each step of a KMC trajectory a large number of such first-principles rate constants is at present and for the foreseeable future in general not feasible. There are currently a number of routes pursued to overcome this showstopper. One obvious remedy would be to recycle first-principles rate constants that have already been computed at previous KMC trajectory steps and thus build up an increasing database. In practice, this requires an unambiguous recognition scheme though that allows to identify that elementary processes that are possible at the present KMC step are identical to processes that were possible in previous steps. Another possibility is to make the rate constant calculations less demanding. This could either be done by resorting to lower levels of theory like using appropriate interatomic potentials instead of DFT or by using more approximate activation barriers \eg from Br\o{}nsted-Evans-Polanyi relationships ({\em vide infra}) that circumvent the costly determination of the transition state. The crucial issue here is always whether then sufficient accuracy is retained to maintain the desired predictive power. This applies most prominently to surface catalysis, where activity and even worse selectivity are highly sensitive to small changes in activation barriers. Finally, one could consider only selective parts of all possible elementary processes, for instance in catalysis by focusing on certain reaction mechanisms only. If such considerations are based on reliable insight (\eg from experimental evidence or other simulations), this can be very elegant. The grain of salt here is that KMC simulations are often employed precisely to find out which of all possible elementary processes crucially govern the statistical interplay. In other words, the objective of KMC is to identify the important parts of process space rather than to assume them in the first place.

In this situation, a prevalent school of KMC implementations resolves the problem by exploiting a crystalline order in the studied system. Under such order, it is possible to map the relevant PES minima onto some suitable form of periodic lattice. Different system states, for instance in a surface catalysis KMC simulation, differ then only in their distribution of adsorbates on the various lattice positions. This type of KMC is referred to as \emph{lattice} KMC. Let us illustrate the idea with the simple example of a surface process such as the diffusion of an adatom on an fcc(100) surface. If the stable PES minima correspond to the fourfold hollow sites, we can immediately establish a lattice model where we include only diffusion processes that allow the adatom to start and end up in one of the (hollow) lattice sites. For each state of the system it is usually necessary to relax the geometric structure, and the atoms may not be exactly on the lattice positions after that. However, one has to choose the lattice in such a way so that atoms at least end up close enough to the lattice positions in order to allow for an unambiguous assignment to one lattice site. This already significantly reduces (and ensures the finiteness of) the number of possible processes, which however remains rather large.

\begin{figure*}
\centering
\includegraphics[width=0.7\textwidth]{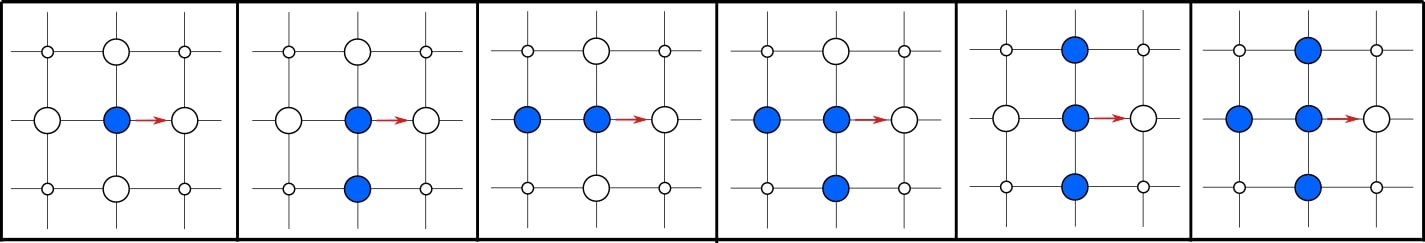}
\caption[]{Hopping diffusion process (red arrow) of a particle on a quadratic lattice for all possible configurations of the nearest neighbor lattice sites in the initial state. Configurations that are equivalent by symmetry are shown only once. The site where the particle diffuses to must be empty for the process to be possible.}
\label{fig:lat_int}
\end{figure*}

Further reduction of the number of required rate constants can be achieved by exploiting the translational symmetry of a crystalline lattice. Let us first assume that in our diffusion toy system there is only one single adatom. The translational symmetry of the crystalline lattice then tells that the elementary processes out of any hollow site are all the same. Once we have computed the rate constants for these processes once, we can simply reuse them for the elementary processes out of another hollow site. They will be the same. In the example of our toy system and if we assume that hopping diffusion over a bridge site is the only diffusion mechanism possible, then this means that we have to simply compute just one single rate constant (for such a bridge hopping process). Here, the symmetry refers to the symmetry of the lattice positions {\em per se}, since (remember) we considered only one adatom in our toy diffusion example. If we have other species occupying nearby lattice positions, this symmetry will readily be broken. What helps in this case is the nearsightedness of chemical interactions as formulated by Walter Kohn. The interaction to some nearby species on the lattice may thus already be so small, that it has a negligible influence on the rate constant. If we neglect such influences between neighboring species to an increasing degree, we restore a complete locality of the elementary processes to the level that we have an as simple and high-symmetry situation as we had in the case of the isolated adatom. For this particular example, the most local approximation would for instance be to completely neglect any interaction with nearby other species -- except for preventing diffusion processes in which an adatom would end up in an already occupied site (so-called site blocking). Despite the enormous number of possible configurations of a system with an arbitrary number of adatoms on the lattice, still only one single rate constant for hopping processes over bridge sites would be required. If we consider a $(10 \times 10)$ lattice, then the state-to-state matrix $k_{ij}$ in Eq.~(\ref{eqn:master}) is still of the order $2^{100}$. However, it only contains elements that are either zero or have this single rate constant as their value. More generally, translational (and maybe any rotational) symmetry does thus not reduce the state space. However, it can dramatically reduce the number of inequivalent rate constants that need to be computed.

As a less drastic approximation we may consider only the lateral interactions of the adsorbate with species in nearest neighbor sites. As illustrated in Fig.\ \ref{fig:lat_int} in our simple diffusion example we will then need to compute five distinct rate constants, one for each possible occupation of the four nearest neighboring sites: one with no nearest neighbors, two with one neighboring site occupied, one with two and and one with three neighboring sites occupied. The generalization to different lattices is a straightforward exercise. Augmenting the local environment that would affect the rate constants would improve the accuracy of the simulation (if such further reaching lateral interactions are indeed still non-negligible), but obviously requires even more rate constant computations. In practice, this gradual inclusion of lateral interactions on a lattice proceeds through cluster-expansion techniques, which we will further discuss in Sec.\ \ref{sec:lat_int}. Essentially, cluster expansions then allow to modify the fundamental rate constant of a given elementary process in the absence of other species to the rate constant of the same process with any distribution of nearby neighbors.

In the lattice approximation one can therefore scan all possible configurations and transitions, compute the associated rate constants for any local configuration beforehand and save the latter in a so-called \emph{rate catalog}. During the KMC run, the current configuration at a trajectory step is examined and the possible processes and their rate constants are extracted from the rate catalog. An alternative is to identify the possible processes and calculate the cluster-expansion correction to the fundamental process rate constants only \emph{on-the-fly} at each KMC step. This can be numerically more efficient when many lateral interactions are taken into account, and the size and cost of searching a comprehensive rate catalog becomes intractable. Examples of this will be provided in Sec.\ \ref{sec:lat_int}.

The approximations used so far can be extremely efficient and may allow for the possibility of using very large supercells in the performed simulations. However, detailed PES information is required and an ordered lattice is assumed as the structural motif. If the lattice model is not suitably chosen, it may neglect important minima of the PES. Simultaneously, any changes of the lattice induced by the simulated dynamical processes cannot be captured by construction \cite{Reuter2016}. This includes important aspects such as reaction-induced surface reconstruction, other surface morphological transitions or loading-induced lattice transformations in intercalation diffusion. The purpose of \emph{off-lattice} KMC is precisely to overcome such limitations and we will discuss in Sec.\ \ref{sec:processes} how the number of rate constant calculations can then be dealt with.\\

\subsection{Mean-field approximation}

An alternative to the full numerical solution of the Master equation (Eq.\ \ref{eqn:master}) with KMC is to introduce further approximations (on top of the lattice approximation) that make the equation easier to solve or allow even for an analytical solution. The most common of such approaches is the mean-field approximation (MFA), where the detailed spatial resolution over the extended lattice is sacrificed and replaced by the mean coverage of each considered species at any of the site types exhibited by the lattice. Mathematically speaking, the MFA assumes that the occupation of the different sites on the lattice is statistically independent, \ie that there are no correlations between different sites on the lattice. In the context of surface adlayers, one says that the adlayer is well mixed.

Let us begin from the (time-dependent) rate $r_{ij}(t)$ of an elementary process, which is given by
\begin{equation}
r_{ij}(t) = P_i(t) k_{ij} \quad .\\
\end{equation}
For a start, let us consider only first- and second-order processes, \ie we assume that at most two lattice sites are involved in the elementary process. Per definition, first-order processes do not involve more than one lattice site, \ie the assumption of uncorrelated lattice sites holds trivially. As an example of a second-order process, let us consider the reaction of two neighboring species $A$ and $B$. We will denote the (time-dependent) pair-probability of finding species $A$ at site $a$ and species $B$ at a neighboring site $b$ with $P_{ab}(A,B,t)$. In the absence of correlations, and assuming that the distribution of the species on the lattice is thus spatially homogeneous, we can write this pair probability as a simple product
\begin{equation}
P_{ab}(A,B,t) = \Theta_a(A,t) \Theta_b(B,t) \quad ,\\
\end{equation}
where $\Theta_a(A,t)$ and $\Theta_b(B,t)$ are the (time-dependent) spatially averaged coverages of species $A$ at sites of type $a$ and species $B$ at sites of type $b$, respectively. Generalizing this to any reaction order, the MFA hereby condenses the high-dimensional Master equation into much simpler rate equations of the form
\begin{equation}
r_{ij}(t) = N_{ij} k_{ij} \prod_{a \in i} \Theta_a(A,t) \quad ,\\
\end{equation}
where $N_{ij}$ is a geometrical factor accounting for the connectivity of the sites involved in the initial and final states $i$ and $j$ and the species $A$ occupies the site $a$ in the initial state $i$ \cite{Matera2011}.

The MFA thus yields a set of coupled differential equations, which are solvable by standard algorithms \cite{Medford2015}. In catalysis, this is often combined with certain assumptions about the rate-determining step (see Sec.\ \ref{sec:sensitivity}) to arrive even at an analytical solution for the reaction rate. While these approximations and the MFA approximation itself simplify the problem at hand enormously, one should always keep in mind that they do represent approximations. In particular the MFA is in general only fulfilled for infinitely fast diffusion and in the complete absence of lateral interactions. We will come back to this point in Sec.\ \ref{sec:timescale}. The first-principles input required for a MFA model is largely the same as for a KMC model, that is, all possible processes and their associated rate constants. However, due to the assumption of (infinitely) fast diffusion inherent to the MFA model, kinetic barriers for diffusion between sites of identical type do not need to be explicitly calculated.\\

\subsection{Codes}

Even though in particular in the context of surface catalysis MFA microkinetic models are still prevalent, one can clearly discern a trend toward KMC simulations. In one way or the other, it is often said that the maturity of a new simulation technique can be judged by the emergence of general-purpose software packages that are user friendly (maybe even up to providing a graphical user interface). If one takes the latter argument at face value, KMC has indeed matured dramatically in the last couple of years. A number of high-end KMC codes have appeared in an astonishingly short period of time. Referring for a more detailed (and maybe exhaustive) account of such codes to the review by Stamatakis {\em et al.} \cite{Stamatakis2014}, we here compile only a brief presentation of a number of such codes to provide an impression of what is presently available:

One of the first general-purpose KMC implementations was provided by Lukkien {\em et al.} in the code {\tt CARLOS} \cite{Gelten1998}. In {\tt CARLOS} one can specify any kind of reaction as input, then the program uses pattern recognition to identify possible reactions. {\tt CARLOS} also implements time-dependent rate constants.

{\tt SPPARKS}, developed by Plimpton {\em et al.} \cite{Slepoy2008}, implements several KMC solvers and is structured modularly to facilitate expansion and implementation of new models and solvers. Currently implemented models include both lattice and off-lattice applications, as well as a general purpose model for the simulation of biochemical reaction networks.

Stamatakis and Vlachos \cite{Stamatakis2011} developed an approach that employs graph-theoretical ideas to overcome the limiting assumption that each participating species occupies a single site and that elementary events involve a maximum of two sites. Here, lattice structure and elementary events are represented as graphs, and lattice processes are identified by solving subgraph isomorphism problems during the simulation.
Building on the latter code, Nielsen, Stamatakis {\em et al.} \cite{Nielsen2013} developed {\tt ZACROS}, which incorporates cluster expansion Hamiltonians in order to accurately account for long-range lateral interactions. The latter two approaches are suitable for treating systems with rather complex surface chemistry, including organic adlayers and more generally situations where the reactants may adsorb on the surface on multiple sites. 

The recently developed MonteCoffee \cite{Jorgensen2018mc} exploits similar ideas to the graph-theoretical approach, geared towards the simulation of nanoparticles. The code uses neighbor lists to represent the site connectivity, rather than mapping the problem onto a lattice. With respect to the graph-theoretical approaches, hereby the user directly controls the site connectivity. 

Adaptive KMC (aKMC) approaches ({\em vide infra}) were mostly developed by Henkelmann and coworkers \cite{Henkelman2001,Xu2008}. The code {\tt EON} currently includes a set of algorithms to model mesoscale dynamics (parallel replica dynamics, hyperdyamics, and basin hopping as well as aKMC). For aKMC, the code implements a server-client architecture where the client processes are responsible for saddle point searches (see Sec.\ \ref{sec:tssearch}) to escape the current basin, then report the calculated rates back to the server which executes the KMC algorithm.

{\tt kmos} \cite{Hoffmann2014b} is an application programming interface based KMC framework that facilitates the generation of an abstract model definition in Python, which is then used to automatically generate efficient Fortran code. The code was instigated by Max Hoffmann and is mainly being developed in our group. We will use it for the hands-on examples in the later sections, providing a detailed practical account on how to run KMC simulations. In the following section we will further describe the KMC algorithm underlying {\tt kmos} as one example of present-day codes, while in general we emphasize that different codes may implement different algorithms to solve the Master equation.

\section{Getting practical: Algorithms and input data}
\label{algo}

As discussed above, the real trick of KMC is the KMC algorithm that generates stochastic trajectories in such a way that their appropriate averaging yields the time evolution of the probability $P_i(t)$ in the Master Eq.~(\ref{eqn:master}). One of the most commonly used such KMC algorithm, initially developed by Bortz, Kalos and Lebowitz \cite{Bortz1975} for Ising spin systems, is known as the BKL algorithm (after the authors) or the "n-fold way". It also goes under the names of Variable Step Size Method \cite{Jansen1995} or Direct Method \cite{Gillespie1976}. For a simple, practical rationalization of the algorithm, let us consider a toy system which can assume only two states $A$ and $B$ connected by a barrier with associated rate constants $k_{AB}$ and $k_{BA}$ for the forward and backward transitions, respectively. In this system, only two elementary events are possible; if, say, the system is sitting in state $A$, it can hop to state $B$. As the rate constant for this hop is $k_{AB}$ (with unit time$^{-1}$), one may naively think that the average time that will have passed until such a hop occurs is $\Delta t_{AB} = k_{AB}^{-1}$. Obviously for the hop back from state $B$ to state $A$, the average time would be $\Delta t_{BA} = k_{BA}^{-1}$. Accordingly, a KMC algorithm would generate a trajectory where after each hop the time is incremented by $\Delta t_{AB}$ or $\Delta t_{BA}$ (depending on what hop occurred). 

Mathematically, this naive thinking is not entirely correct. In reality, while being in state $A$ for each short increment of time the system will have the same probability of finding the escape path. This generates an exponentially decaying survival statistics, whose derivative represents the probability distribution $p_{AB}$ for the true time of first escape:
\begin{equation}
p_{AB}(t) = k_{AB} \exp(-k_{AB}t) \quad .
\end{equation}
The average escape time thus has to be appropriately weighted by this Poisson distribution. It can be shown \cite{Gillespie1976, Fichthorn1991} that this is achieved by advancing the system clock by
\begin{equation}
\label{eqn:escapeAB}
 \Delta t_{AB} = - \frac{{\rm ln}(\rho_2)}{k_{AB}} \quad ,
\end{equation}
where $\rho_2 \in]0,1]$ is a randomly drawn number. 

When we now generalize this to an arbitrary system, where at each step along a KMC trajectory a multitude of processes $i \rightarrow j$ 
from the current state $i$ to other states $j$ are possible, the theory of Poisson processes allows to straightforwardly derive the average time that will have passed until {\em any} process has occurred. Since the elementary processes are independent, each has its own probability distribution $p_{ij}$ given by
\begin{equation}
p_{ij}(t) = k_{ij} \exp(-k_{ij}t) \quad ,
\end{equation}
and the probability of the time of first escape from state $i$ through \emph{any} of the processes $i \rightarrow j$ follows as 
\begin{equation}
p_{\rm escape}(t) = k_{\rm tot} \exp(-k_{\rm tot}t) \quad ,
\end{equation}
where 
\begin{equation}
\label{eq:k_tot}
k_{\rm tot} = \sum_{j}  k_{ij}
\end{equation}
is the total escape rate constant obtained as the sum of all the individual elementary rate constants. Just as for the single process case, the average escape time is then
\begin{equation}
\label{eqn:escape}
 \Delta t_{\rm escape} = - \frac{{\rm ln}(\rho_2)}{k_{\rm tot}} \quad ,
\end{equation}
where $\rho_2 \in]0,1]$ is a randomly drawn number. It is crucial to highlight that this escape time only depends on the total rate constant, and is independent of the actual process that actually occurs to bring the system out of the current state $i$. This actual process nevertheless needs to be identified, since this is what determines to which state $j$ the system propagates. This state $j$ is then the starting point for the next KMC step, \ie the next KMC step evaluates the escape from this particular state $j$. 

\begin{figure*}
\centering
\includegraphics[width=1.0\textwidth]{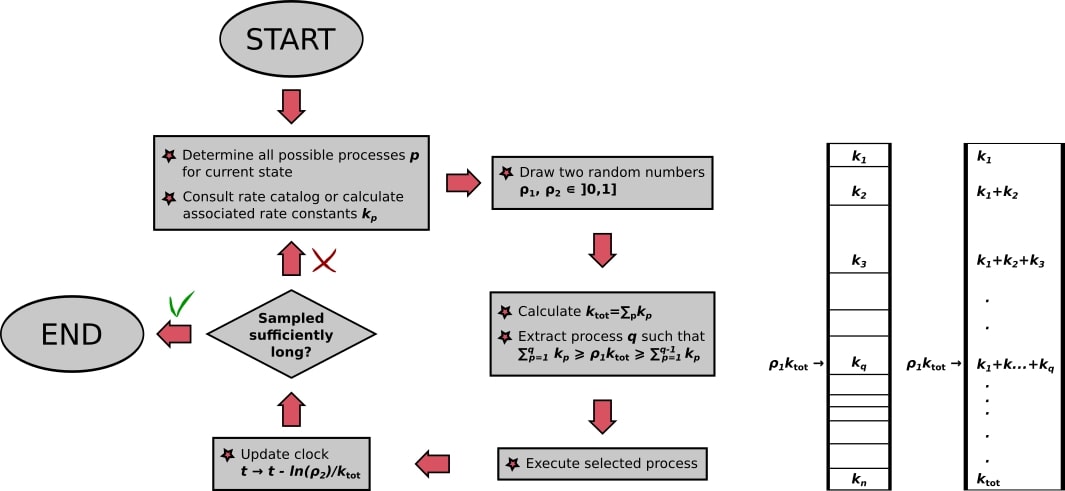}
\caption[]{Left: Flowchart of the BKL algorithm as implemented in {\tt kmos}. Right: Graphical illustration of random process selection in the process stack.}
\label{fig:kmc_flow}
\end{figure*}

The BKL algorithm determines this executed process out of all possible processes again by rolling the dice. Imagine a stack of segments of height proportional to the rate constants $k_{ij}$ of the possible processes, which correspondingly sums up to a total height of $k_{\rm tot}$. We can choose a process to execute out of this stack by drawing a random number $\rho_1$ and multiplying it by $k_{\rm tot}$. The resulting number will ``point'' at the process with correct probability in the process stack, as illustrated in Fig.\ \ref{fig:kmc_flow}. The selected process is then executed, bringing our system into a different state for the next KMC step. This way of choosing out of the stack ensures that faster processes are selected with a higher probability than slower ones: They have a larger rate constant, have correspondingly a thicker segment in the stack and are correspondingly chosen more often. We therefore have a recipe to generate trajectories that satisfy the Master equation, obtained stochastically via the extraction of only two random numbers $\rho_1, \rho_2 \in]0,1]$. An important parameter is thereby also the random number seed used to generate the sequence of random numbers, as recalculating the trajectory with a new seed value will lead to the new trajectory that is statistically independent of the former.

The algorithm, also illustrated in Fig.\ \ref{fig:kmc_flow}, may thus be summarized as follows:
\begin{itemize}
 \item Make a list of possible processes $p$ for the system to escape the current state $i$, with associated rate constants $k_p$;
 \item draw two random numbers $\rho_1, \rho_2 \in]0,1]$;
 \item calculate $k_{\rm tot}=\sum_p k_p$;
 \item extract process $q$, which has to fulfill the constraint $\sum_{p=1}^q k_p \geqslant \rho_1 k_{\rm tot} \geqslant \sum_{p=1}^{q-1} k_p$;
 \item execute randomly drawn process;
 \item update clock: $t \rightarrow t - {\rm ln}(\rho_2)/k_{\rm tot}$.
\end{itemize}

\section{Rate constants from first principles: Transition state theory} 
\label{sec:tssearch}

As discussed in the previous sections, KMC requires rate constants for all considered elementary processes as an input. For the surface applications focused on here, these rate constants are at present predominantly obtained through Transition State Theory (TST). Making a number of assumptions, for instance that the flux of trajectories passing the transition state (TS) from state $i$ to $j$ will never come back to the state $i$ (no-recrossing rule) and that the barrier crossing is a purely classical event (no tunneling), TST provides a simple expression for the rate constant, known as Eyring (or Eyring-Polanyi) equation: \cite{Laidler1987}
\begin{equation}
\label{eq:tst}
\begin{split}
k^\mathrm{TST}_{ij} = &\ \frac{q^{\mathrm{vib}}_{\mathrm{TS}}}{q^{\mathrm{vib}}_{i}} \frac{k_{\rm B} T}{h} \exp\left(-\frac{\Delta E_{ij}}{k_{\rm B} T} \right) \\
= &\ k_o \frac{k_{\rm B} T}{h} \exp\left(-\frac{\Delta E_{ij}}{k_{\rm B} T} \right) \quad ,
\end{split}
\end{equation}
where $T$ is the absolute temperature, $h$ is Planck's constant, $q^{\mathrm{vib}}_{\mathrm{TS}}$ and $q^{\mathrm{vib}}_{i}$ are the partition functions at the transition state and at the initial state, respectively, and $\Delta E_{ij}$ is the activation barrier of the process. The latter is directly available from PES information and therefore accessible to first-principles calculation (\eg semi-local DFT). The prefactor $k_o \frac{k_{\rm B} T}{h}$ may in principle be calculated. Most popular is harmonic TST, where the partition functions are obtained from the vibrational modes at the initial state $i$ and at the TS. In the predominant number of studies the considerable computational costs of these vibrational calculations are avoided though, and one simply approximates $k_o \simeq 1-10$, yielding a prefactor in the range $10^{12}-10^{13}~\mathrm{s}^{-1}$. This is never really fully justified, and if at all only when the vibrational properties of the TS do not differ much from those of the initial state. A prominent class of processes where this common approximation is not valid are non-activated adsorption processes, where the prefactor needs to account for the strong entropy reduction from the gas phase to the surface-bound state. In this case the rate constant is better estimated as: \cite{Reuter2006,Chorkendorff}
\begin{equation}
\label{eq:unactivated_ads}
k^{\rm ads}_{n,B} (T, p_n) = \widetilde{S}_{n,B}(T) \frac{p_n A_{\rm uc}}{\sqrt{2\pi m_n k_{\rm B} T}} \quad ,
\end{equation}
where $p_n$ is the partial pressure of species $n$ of mass $m$, and the local sticking coefficient $\widetilde{S}_{n,B}(T)$ governs the fraction of impinging particles sticking to a site $B$ located in a surface unit cell of area $A_{\rm uc}$.\\

\subsection{Master equation and detailed balance}

We will next motivate some practical guidelines that the input (processes and rate constants) to any microkinetic model (KMC or MFA) must adhere to. Let us consider a system that has reached steady state. This imposes the constraint of vanishing derivative in the Master equation (Eq.\ \ref{eqn:master}), which leads to the condition
\begin{equation}
\label{eq:detbal1}
\sum_{j \neq i} \left[ k_{ij}P^*_i - k_{ji}P^*_j\right] = 0 \quad ,
\end{equation}
where $P^*_{i}$ ($P^*_{j}$) is the time-independent probability that the system is in state $i$ ($j$). This condition is a conservation law stating that the sum of the rates of all transitions out of any state $i$ $(j)$ must equal the sum of the rates of all transitions into state $i$ $(j)$. At thermodynamic equilibrium, microscopic reversibility and the principle of detailed balance \cite{Tolman1925} imposes the even stronger constraint that the average rate of every microscopic process must exactly balance its reverse process
\begin{equation}
\label{eq:detbal2}
\frac{k_{ij}}{k_{ji}} = \frac{P^*_j}{P^*_i} \quad .
\end{equation}
The right-hand side of Eq.\ \ref{eq:detbal2} is thereby proportional to the states' Boltzmann weights and can thus be expressed in terms of the free energy difference between states $i$ and $j$:
\begin{equation}
 \label{eq:free}
 \frac{k_{ij}}{k_{ji}} = \exp \left( -\frac{F_j(T)-F_i(T)}{k_{\rm B} T} \right) \quad . 
\end{equation}

The above derivation motivates two practical guidelines for constructing KMC (or MFA) models. The first guideline is that every microscopic process must have defined a corresponding reverse process, and the second guideline is that the rate constant expressions used for the forward and reverse processes must fulfill Eq.\ \ref{eq:free}.

For the latter point it is particularly crucial to realize that this also extends to computing the free energies of both states $F_i$ and $F_j$ with the same numerical approximations. In practice, this is often neglected (different supercells/configurations, mix of first-principles and empirical data etc.) and can then have drastic consequences as the kinetic model is not thermodynamically consistent \cite{Mhadeshwar2003,Schmitz2000}. Quite some work has therefore been devoted to achieve an overall thermodynamic consistency, \eg \cite{Mhadeshwar2003,Nielsen2013}.\\

\subsection{Obtaining rate constants: Transition state search}

In the context of determining rate constants, it is natural to put a primary focus on the \emph{lowest} activation barriers that need to be overcome, \eg in catalysis on the \emph{minimum energy path} connecting reactants and products. From the mathematical standpoint, locating the lowest barrier(s) translates into locating the lowest first-order saddle point(s) on the PES, which is  a particularly challenging task for which -- in contrast to locating PES minima -- there is yet no general approach that is guaranteed to work. In the following we will briefly discuss the reliability, accuracy and performance of different available methods. Always keeping in mind that ``your mileage may vary'', there is nevertheless a general guideline as to which family of approaches would be preferable depending on the nature of the problem at hand. For lattice KMC one would assume or derive the mechanism (\eg the reaction mechanism in the context of heterogeneous catalysis) and subsequently compile a list of the elementary processes that constitute this mechanism. In this case, initial and final states are therefore pre-determined, enabling the use of so-called {\em interpolation} methods. For adaptive KMC, more often than not no previous assumption of mechanism is made and one may need to blindly explore the possible and most probable escape pathways from a current state. So-called {\em local} methods are then mandatory, as only information on the initial state is available.\\

\subsubsection{Interpolation methods}
\label{sec:ts_interp}

The simplest form of interpolation-based TS search consists in identifying a reaction coordinate guess in one or a small number of internal degrees of freedom, preferring those that describe the main structural differences between initial and final state. The selected coordinates are subsequently constrained to specific values between the initial and final structures, while all remaining degrees of freedom are optimized. Such TS searches are often referred to as \emph{coordinate-driving} (\emph{drag}) methods \cite{Halgren1977,Rothman1980,Williams1982}. The success of drag methods depends critically on the ability to choose a good set of reaction coordinates and on the topology of the PES in the direction of the remaining degrees of freedom. In general, if the reaction path is dominated by only one or two degrees of freedom, the coordinate driving may work, and the constrained optimized geometry (with the smallest residual gradient) is a good approximation to the TS. On the other hand, a bad choice of reaction variable(s) may lead to hysteresis and converge to (unphysical) discontinuous reaction paths \cite{Halgren1977,Henkelman2002}.

Drag methods operate on one structure at the time. A significant improvement is achieved by simultaneously optimizing multiple points along the initial guess of the reaction path. An example is the \emph{ridge} method \cite{Ionova1993}, which iteratively refines an initial guess of the TS by simultaneously relaxing two replicas of the latter, sligthly displaced across the ridge of the PES, until they contract to the actual TS. Methods that operate with more than two structures are often referred to as ``chain-of-state'' methods. The initial distribution of structures will typically be along a linear interpolation of coordinates between the initial and final, or any convenient form of continuous variation along a chosen reaction path. All intermediate states or images are then optimized simultaneously in some concerted way, providing not only the saddle point, but also a good approximation of the entire reaction path.

\begin{figure*}
\centering
\includegraphics[width=1.0\textwidth]{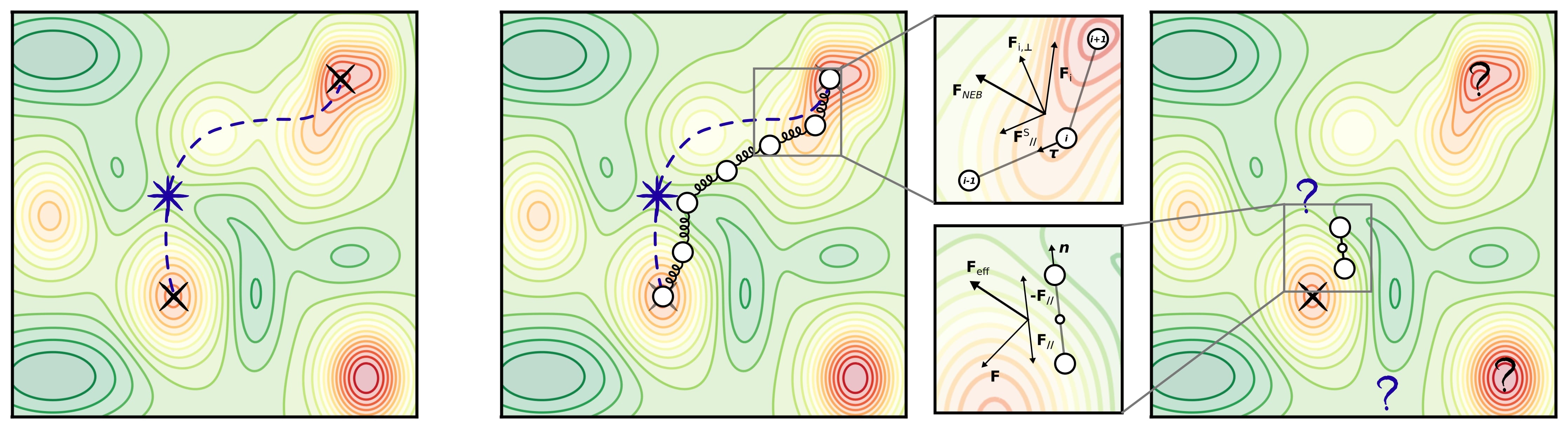}
\caption[]{Transition states and most popular transition state (TS) search methods. Left: An arbitrary PES exhibiting multiple minima (indicated by red colors). Black crosses mark two local minima, \eg initial state (IS) and final state (FS) of a known elementary reaction. A violet star marks the location of the TS; a violet dashed line represents the minimum energy path connecting IS and FS. Center: Illustration of the NEB method. The magnified panel shows the composition of forces which determines the effective NEB force acting on an image $i$ at a certain optimization step. Right: Illustration of the dimer method. The magnified panel shows the composition of forces which determines the effective force acting on the dimer at a certain optimization step, after the constrained minimization with respect to orientation.}
\label{fig:pes_tst}
\end{figure*}

Among those, the \emph{Nudged Elastic Band} (NEB) method \cite{JNSSON1998,Henkelman2000} is arguably most popular as it incorporates strong points of older approaches in order to cure their shortcomings. After initializing an initial chain of images $\mathbf{R}_i$, NEB minimizes a target function (``elastic band'') defined as the sum of energies of all intermediate images and an additional penalty term which distributes the points along the path through a single spring constant $k$ (see Fig.\ \ref{fig:pes_tst}):
\begin{equation} \label{eq:NEB}
S_{\mathrm{EB}}(\mathbf{R}_1,\ldots, \mathbf{R}_N) = \sum_{i=1}^{N} E(\mathbf{R}_i) + \sum_{i=1}^{N-1} \frac{1}{2}k(\mathbf{R}_{i+1}-\mathbf{R}_{i})^2 \quad .
\end{equation}
In general, a straightforward minimization of $S_{\mathrm{EB}}$ would exhibit a tendency to ``cutting corners'' if the spring constant $k$ is too large, and ``sliding down'' towards the extrema if $k$ is too small (thus undersampling the actual TS region). These problems are alleviated by ``nudging'' the elastic band, \ie by using only the component of the spring force parallel to the tangent of the path (to cure for corner-cutting), and only the perpendicular component of the energy force (to cure for down-sliding). The total force acting on each image is then
\begin{equation} \label{eq:NEBforce}
\begin{split}
\mathbf{F}_{i,{\rm NEB}} = &\ \mathbf{F}^s_{i\parallel} - \mathbf{F}^i_{\perp} \\
= &\ k \left( |\mathbf{R}_{i+1} - \mathbf{R}_i | - | \mathbf{R}_i - \mathbf{R}_{i-1} | \right ) \hat{\tau}_{i}  - \nabla E(\mathbf{R}_i)_{\perp} \quad ,
\end{split}
\end{equation}
$\hat{\tau}_{i}$ being the tangent unit vector at image $i$. In the Climbing-Image NEB (CI-NEB) \cite{Henkelman2000_ci} variant, the image with the highest energy is selected after a few iterations and driven up towards the saddle point by turning off its spring force and reversing the component of the potential force parallel to the chain. This yields exactly the saddle point (which in the non-climbing version is obtained by interpolation) at no additional computational cost.

The NEB method is still likely to run into trouble when dealing with a PES for which the energy varies largely along the reaction path, but very little perpendicular to it. Regarding this problem, it has been pointed out that for CI-NEB corner cutting does not affect the accuracy of the TS, and that a more robust relaxation to the TS may be achieved by using the full spring force rather than only the component parallel to the tangent of the path \cite{Kolsbjerg2016}. Furthermore, as all such algorithms in its category, NEB comes at a high computational cost as it involves the optimization of many structures and typically requires a rather large number of iterations. Technical parameters such as the number of images and the value of the spring constant must be wisely chosen beforehand. The latter issue in particular may prove to crucially influence the optimization efficiency: a small value causes an erratic coverage of the reaction path, while a large value focuses the effort on distributing the points rather than on finding the reaction path, and consequently slows down the convergence. In the traditional NEB method the number of images are fixed during the simulation and it can be challenging to reach a good compromise between a sufficient coverage of the reaction path and the computational cost. To mitigate this, an automated NEB (AutoNEB) algorithm has been proposed \cite{Kolsbjerg2016}, which can save computational resources by focusing first on converging a rough path before improving on the resolution around the TS. An alternative solution to the same problem has been proposed in the truncation-based energy weighting string method \cite{Carilli2015}, which uses energy weighting to focus the computational effort on the physically interesting images within the barrier region. Finally, it is also worth mentioning the growing string method (GSM) \cite{Peters2004}, which circumvents the need for a (good!) initial guess of the reaction path by separately evolving two string fragments, one associated with the reactants and the other with the products, until the fragments converge and thereby define the reaction path. The combination of GSM with an eigenvector following TS search \cite{Zimmerman2013} has shown promising results for a benchmark set of more than 100 elementary reactions.\\

\subsubsection{Local methods}
\label{sec:ts_loc}

Keeping in mind that transition states are points where the gradient vanishes, they may in principle be located by minimizing the gradient norm. This is exactly the working concept behind the so-called local methods. In contrast to interpolation methods, local methods only use information of the PES function and its first and possibly also second derivatives at each point, \ie they require \emph{no} knowledge of the initial-state and/or final-state geometries. They do, however, usually require a good estimate of the TS to use as a starting geometry in order to converge. This is one of their main limitations.

A most common member of the local group of methods is the Newton-Raphson (NR) approach, which locates the TS directly, given that one starts the search sufficiently close to the TS. Sufficiently close here means already in the harmonic region with the Hessian having exactly one negative eigenvalue. Under these conditions, computing the Hessian and inverting the second order Taylor expansion directly yields the step which maximizes the energy along the TS eigenvector and minimizes the energy along all other directions, converging exactly to the TS. The main drawback of the NR method is the need for generating and manipulating the full Hessian matrix.

However, the main function of the Hessian for saddle point optimization is to provide the direction along which the energy should be maximized (lowest ascent if sufficiently near the TS). The dimer method \cite{Henkelman1999} can be employed to determine this direction without calculating the Hessian explicitly, employing two symmetrically displaced replicas--the dimer (see Fig.\ \ref{fig:pes_tst}), which is used to transform the force in such a way that optimization leads to convergence at a saddle point rather than at a minimum.
In general, the strategy for the dimer method is to try many different initial configurations around a minimum, usually taking them from the extrema of a short high-temperature MD trajectory in order to find the saddle points that lead out of that basin. In a first step, the dimer is minimized with respect to orientation by imposing a constrained distance between the images. The lowest mode direction is then given by the line connecting the two images, and this can be used for displacing the central structure, \ie translating the dimer, which is then followed by a new dimer optimization and so on. The force acting on the centre of the dimer gets modified by inverting the component in the direction of the dimer: minimization of this force drives the dimer to a saddle point. A dimer optimization can be done using only first derivatives, and thus alleviates the need for calculating the Hessian matrix. In general, however, performance scaling with system size is not really known. It is unclear in particular whether the added computational cost of optimizing each dimer configuration eventually outweighs the saving by not requiring an explicit Hessian.\\

\subsubsection{BEP and scaling relations}
\label{sec:ts_scaling}

An alternative, cheaper approach to determine activation barriers, which still retains grounds in first principles, is provided by the employment of approximate energy relations. The most prominent example is the Br\o{}nsted-Evans-Polanyi (BEP) relation \cite{Norskov2002,Michaelides2003}, which yields linear relationships of the kind
\begin{equation}
 \Delta E \simeq c_1 (E_{\rm FS} - E_{\rm IS}) + c_2 \quad ,
\end{equation}
where $c_1$ and $c_2$ are constants and $E_{\rm FS}$ and $E_{\rm IS}$ are the total energies of the final and initial states, respectively. The latter are obtainable from local geometry relaxations, and hence significantly cheaper than even the sloppiest TS search. The two parameters $c_1$ and $c_2$ need to be determined by linear fitting to appropriate first-principles calculations. The parameters are in general only transferable to site types that are similar to those used in the fitting, \eg the fcc(111) sites of transition metal surfaces, but they can be rather universal among different kinds of reactions \cite{Wang2011a}.

Furthermore, it has been shown that the binding energies of many reactants, products and intermediates at transition metal surfaces correlate with the binding energies of the few base elements (mostly C, N, O, S, halogens) with which the molecules typically bind to the surface \cite{Abild2007}. The employment of such scaling relations, combined with BEP relations, enables an enormous reduction of the computational cost of getting first-principles rate constants for applications such as catalyst screening, where a large number of rate constants have to be computed for ``similar'' processes.

\section{Garbage in -- garbage out}
\label{sec:processes}
As always in modeling, it is important to realize that the predictions made from a model are limited by the quality of the input data to the model, commonly expressed as the \emph{garbage in -- garbage out} (GIGO) principle. The necessary input data to a KMC model are the possible processes as well as their associated rate constants. As discussed above, in the context of first-principles KMC simulations where the rate constants are determined by DFT or other electronic structure calculations, these input data are at present typically determined beforehand. In other words, rather than the KMC simulation identifying by itself what processes are important or should be considered and at which accuracy, the simulation depends on a fixed given list of such processes with rate constants that come with the typical uncertainty imposed by the underlying DFT calculation (the rate catalog). This rigid setup leads to a high sensitivity of the KMC simulation results on this input data. 

As already mentioned initially, advanced KMC approaches that overcome at least some of the limitations of this prevalent rigid input-data setup are a long sought goal and topic of current research. One possibility to automatically identify the relevant processes in a system would be to use (accelerated) MD approaches like hyperdynamics, temperature-accelerated dynamics or replica-exchange dynamics beforehand \cite{Voter2002}. In adaptive (on-the-fly) KMC a process search using the dimer method or high-temperature MD simulations is directly integrated into the KMC algorithm \cite{Henkelman2001,Chill2014}. A great advantage of the latter methods is also that the KMC model does not necessarily have to be implemented on a fixed lattice. However, they are generally also much more computationally demanding, as they require many orders of magnitude more energy and force evaluations to determine all processes and their barriers. Applications have therefore been limited to rather simple systems or systems where the energy and force evaluations can be done with classical force fields instead of DFT \cite{Xu2008,Konwar2011,Pedersen2015}. At least for the time being, first-principles KMC simulations in the application areas covered here are instead in practice only tractable within the rigid setup, which is why we concentrate on it from now on in this practical guide. 

In the following we will illustrate the above-described concepts using a simple model for the diffusion of Au adatoms on a Au(100) surface. This will highlight possible pitfalls, but will also provide guidance for best practice. All discussed KMC models have been implemented in the {\tt kmos} software package \cite{Hoffmann2014b} and are available in the supplemental data.\\

\subsection{Adatom diffusion on Au(100)}
\label{sec:diff_example}

\begin{figure}
\centering
\includegraphics*[width=\columnwidth]{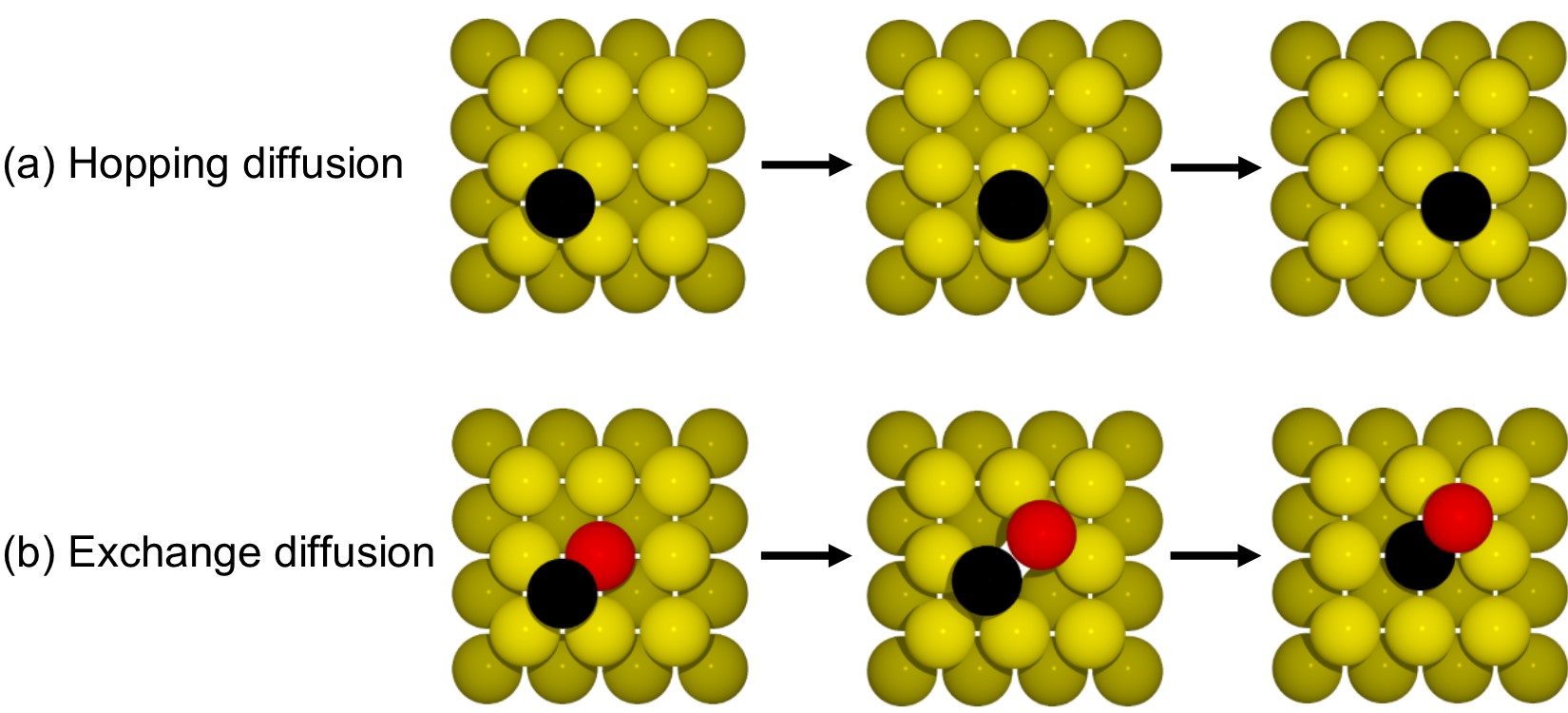}
\caption[]{Illustration of the mechanism for (a) hopping and (b) exchange diffusion (see text).}
\label{fig:diff}
\end{figure}

\begin{figure*}
\centering
\includegraphics[width=.6\textwidth]{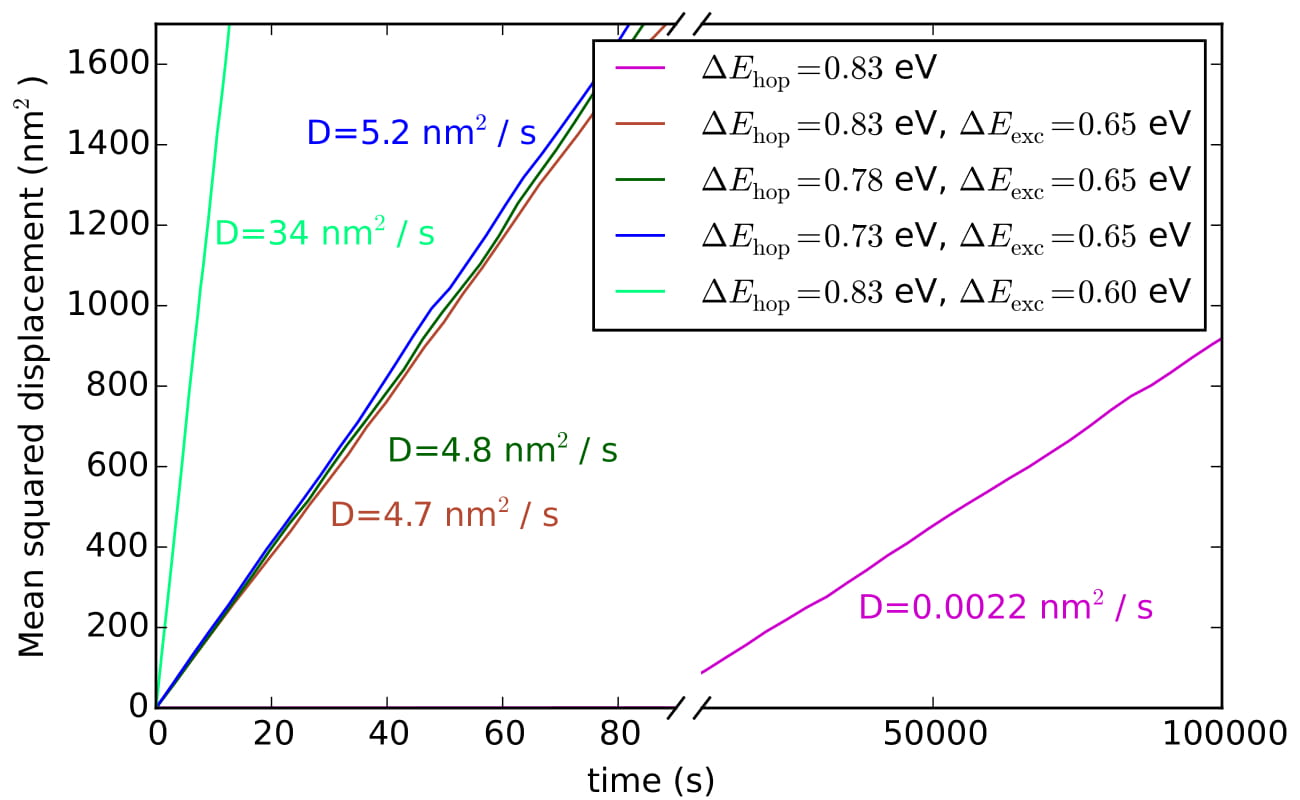}
\caption[]{Averaged mean squared displacements and diffusion constants of adatoms on Au(100) for hopping and exchange diffusion processes and various associated barriers.}
\label{fig:MSD_Au}
\end{figure*}

The simplest diffusion process one can think of in this system consists of the hopping of Au atoms between the fourfold hollow adsorption sites offered by the Au(100) surface lattice, see Fig.\ \ref{fig:diff}(a). For a simple KMC model of this system we will consider a $(20\times20)$ square lattice of adsorption sites. The possible processes are the hops of particles from one site into one of the four neighboring sites up, down, left and right. Neglecting any possible lateral interactions between Au adatoms, the barrier for any of these processes is the same and at the DFT-LDA level it has been calculated to $\Delta E_{\rm diff} = 0.83$\,eV \cite{Yu1997}. We will use the TST expression (Eq.\ \ref{eq:tst}) to express the rate constant of the diffusion processes in terms of this barrier. For simplicity we will ignore any entropic corrections to the barrier and zero-point vibrational energy corrections, \ie $k_o = 1$. In order to run the KMC model we need to fill some of the lattice sites with particles. If we were to initialize the simulation with an empty lattice, we would directly hit a deadlock where no processes are possible. In this example we prepare the initial state by randomly filling sites with Au adatoms corresponding to a coverage of 0.25\,monolayer (ML), \ie every fourth surface site is occupied. As the output of the simulation we calculate the diffusion coefficient of the Au adatoms by tracking the mean squared displacements (MSD) \cite{Garhammer2017} as a function of time. The diffusion coefficient $D$ can then be calculated as
\begin{equation}
D = \frac{\langle {\rm MSD}(t) - {\rm MSD}_0 \rangle}{2 d t} \quad,
\end{equation}
where $t$ is the simulation time and $d$ is the lattice dimension (2 in this case). The averaging $\langle \rangle$ is performed over all Au adatoms. In order to improve the statistics 25 independent simulations were run, which differed from each other in the random initialization of the lattice and in the random number seed used. The pink line in Fig.\ \ref{fig:MSD_Au} shows the average of these 25 simulations from which the diffusion coefficient is determined to be 0.0022 nm$^2$/s.

Such hopping diffusion processes are, however, not the only mechanism possible for self-diffusion on metal surfaces. Both experimentally and theoretically an alternative exchange diffusion mechanism has been described \cite{Bassett1978,Wrigley1980,Feibelman1990,Yu1997}, in which a metal adatom replaces a surface layer atom and pushes it to a neighboring adsorption site, see Fig.\ \ref{fig:diff}(b). With the adatom and surface atom of the same species, this effectively also results in a net displacement. In Ref.\ \cite{Yu1997} it was found that this process occurs with the lower barrier of only 0.65\,eV on Au(100), and should therefore dominate over hopping diffusion. When we additionally allow for this diffusion process in the model, the output of the simulation changes indeed dramatically, since the barrier enters the rate constant for diffusion exponentially. Including exchange diffusion, the diffusion coefficient is now determined to be 4.7 nm$^2$/s, see brown line in Fig.\ \ref{fig:MSD_Au}, \ie more than three orders of magnitude higher than before. It can also be noted that the timescale reached in the simulation (for the same number of total KMC steps) is much shorter with exchange diffusion. In general, the time advanced with a KMC step, $\Delta t_{\rm escape}$, and therefore the total timescale that is computationally reachable during a simulation, depends on the sum of all rate constants (see Eq.\ \ref{eqn:escape}), which is dominated by the fastest process in the system.

The above example thus highlights the extent to which the outcome of a KMC simulation is dependent on knowing and allowing for all relevant processes in the model. This is a severe limitation and requires utmost caution and care in setting up a KMC model. The likelihood to overlook a non-intuitive process such as exchange diffusion cannot be overemphasized and then - alas - the GIGO principle applies with full might. Another drawback of first-principles KMC is that the DFT energies entering the rate constants can have rather large errors associated with them. For processes at extended surfaces often low-rung semi-local DFT functionals (possibly with some $+U$ correction) are still the state-of-the-art. This means that DFT errors on barriers may easily be on the order of 0.1 -- 0.2\,eV. Considering that these barriers enter the rate constants exponentially, see Eq.\ \ref{eq:tst}, the associated rate constants could be wrong by orders of magnitude. However, in general, errors associated with the various rate constants entering a KMC model do not all have a similar effect on the output of the simulation. To illustrate this, we plot in Fig.\ \ref{fig:MSD_Au} the output of the diffusion model including both hopping and exchange diffusion, when lowering the barrier for hopping diffusion by 0.05\,eV (dark green line) and by 0.1\,eV (blue line).
As can be seen, the result only begins to change significantly when the barrier for hopping diffusion comes within few tens of meV of the barrier for exchange diffusion. In contrast, lowering the rate constant for the dominant exchange diffusion process by only 0.05\,eV (light green line) leads to a seven times higher diffusion constant. Again, this is a result of the exponential dependence of the rate constant on the barrier. In other words, a DFT error of 0.1\,eV on hopping diffusion would not change the output of the simulation, while the same DFT error on exchange diffusion would lead to huge changes in the output. 

While this behavior is trivial to guess in this simple two-parameter model, KMC models, particularly in catalysis, can be much more complicated with very many reaction steps and competing pathways. It is then of interest to identify more systematically those processes and their input rate constants that are most important for the outcome of the simulation. Within the more general context of multiscale modeling, viewing first-principles KMC simulations as a hierarchical multiscale modeling setup combining an electronic structure with a mesoscopic statistical technique, such endeavors are called sensitivity analyses. As has hopefully become clear from this simple adatom diffusion model, for the rigid KMC setup such analyses are absolutely pivotal to assess the meaningfulness of the obtained results. In addition, such analyses also provide important insight into which processes are those that control the kinetics, as it is the rate constants of these processes that critically determine the simulation outcome. Such mechanistic insight is another important reason for conducting KMC simulations. In catalysis, these controlling processes are called rate-determining steps, and identifying rather than assuming them for a catalytic cycle is one of the big assets of comprehensive KMC simulations. We will come back to this topic in Sec.\ \ref{sec:sensitivity} after having discussed examples for such more complex catalysis KMC models in the next section.

\section{Steady-state and transient simulations for surface catalysis}
\label{sec:steady_state}

For catalysis applications one is often interested in the behavior of the system once it has reached steady state (see Eq.\ \ref{eq:detbal1}). Since the system is open (constant influx of reactants, constant outflux of products), this still requires kinetic simulations even though the quantities of interest are {\em per se} generally not time-dependent. Major such quantities of interest are the surface composition (for instance in form of averaged coverages of different adsorbates/reaction intermediates), reaction mechanisms and production rates of various chemicals. The latter reaction rate is often expressed as a turnover frequency (TOF), which is the average rate of production of a certain molecule per second per surface site (or surface area). Alternatively, for analysis methods like
temperature programmed reaction \cite{Jansen1995,Rieger2008}, cyclic voltammetry \cite{Rai2006} or titration \cite{Piccinin2014} one might be interested in the transient behavior of a system prepared in a given initial state. In the next section we will introduce a simple catalysis KMC model for CO oxidation on RuO$_2$(110) and illustrate the preparation of various initial states and possible pitfalls with the relaxation to the steady-state solution.\\

\subsection{CO oxidation on RuO$_2$(110)}
\label{sec:cat_example_1}

\begin{figure}
\centering
\includegraphics[width=\columnwidth]{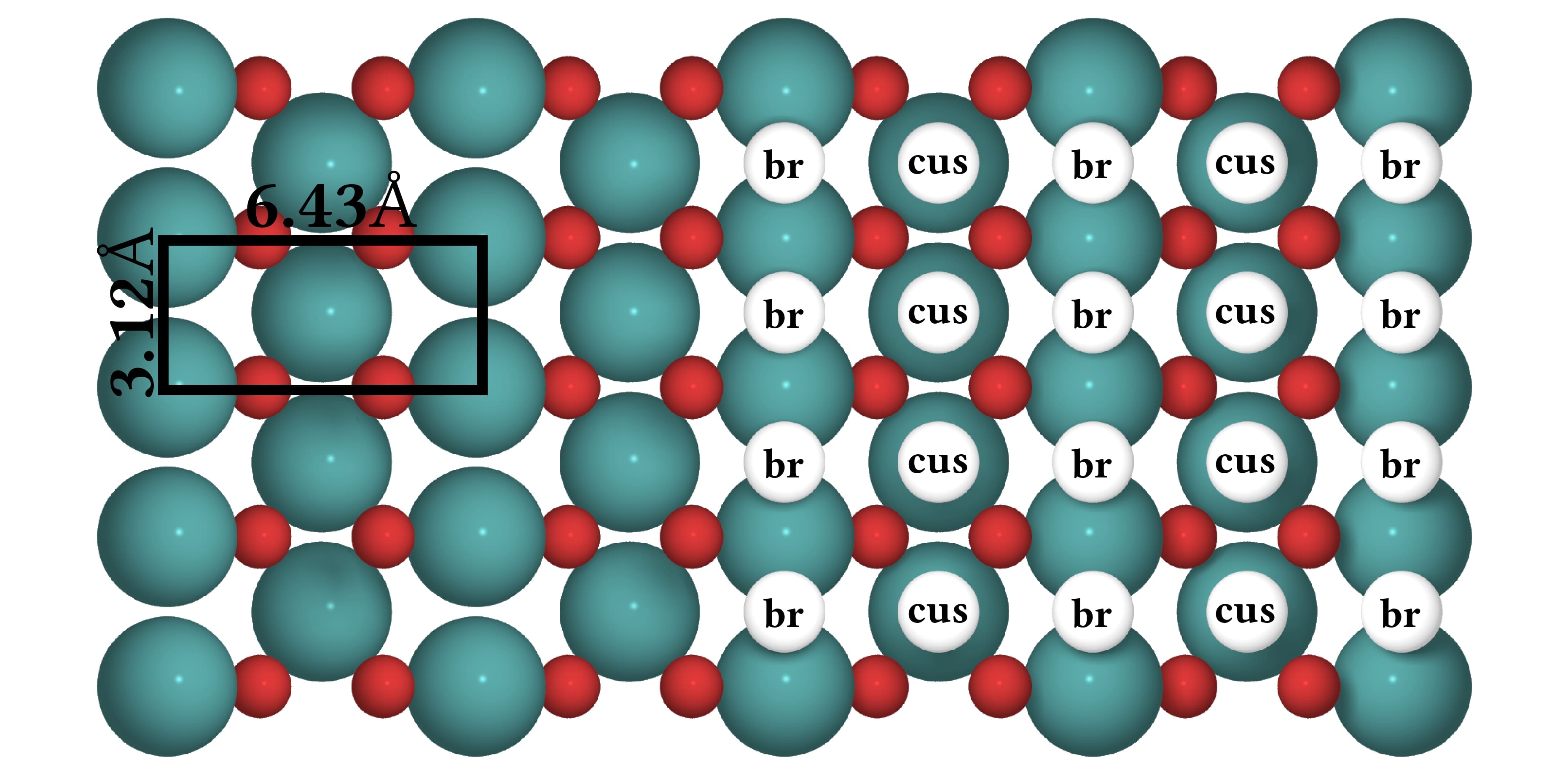}
\caption[]{Top view of the structure of the RuO$_2$(110) surface. To the left, the surface unit cell is shown as a black rectangle. Big green spheres = Ru atoms, small red spheres = O atoms. To the right, the coarse-grained lattice structure is sketched. It consists of alternating columns of cus (coordinately unsaturated) and br (bridge) sites.}
\label{fig:ruo2-lattice}
\end{figure}

The CO oxidation model is taken from Ref.\ \cite{Reuter2004} and we refer to this publication for its more detailed motivation. In this model, the RuO$_2$(110) surface is considered to contain two types of active sites, br (bridge) and cus (coordinately unsaturated) sites, arranged in an alternating rectangular lattice as shown in Fig.\ \ref{fig:ruo2-lattice}. Each site can be either empty or occupied by O or CO. A total of 26 processes are possible in this system, covering non-dissociative CO adsorption/desorption, dissociative O$_2$ adsorption/desorption, diffusion of O and CO, as well as CO$_2$ formation, where for each reaction type all combinations of site types are taken into account. The formed CO$_2$ is assumed to desorb instantaneously and irreversibly due to its weak binding to the surface.

\begin{figure*}
\centering
\includegraphics[width=0.8\textwidth]{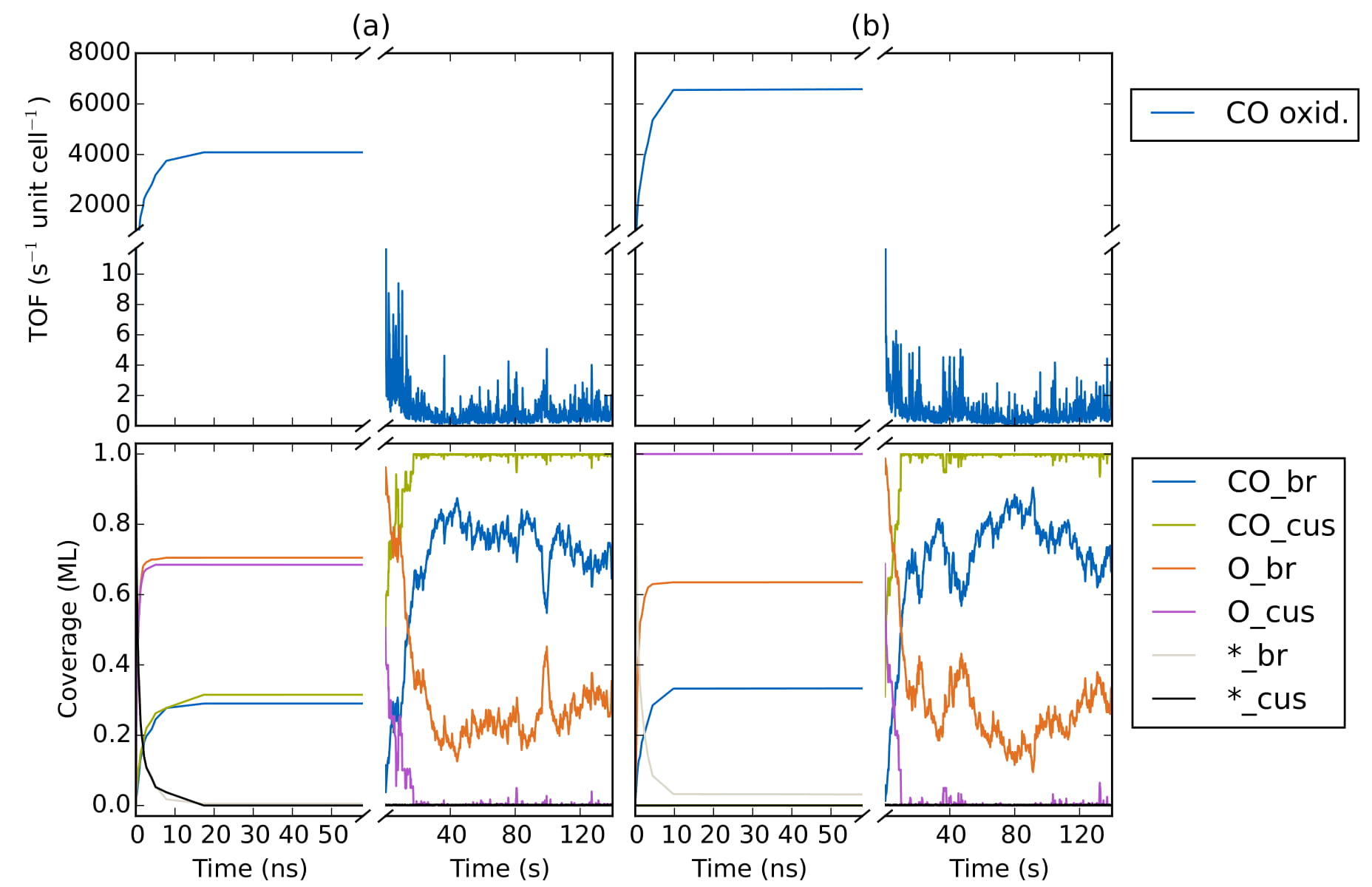}
\caption[]{Temporal evolution of (top panel) CO oxidation TOF and (bottom panel) surface coverage of CO, O and vacancies on the cus and bridge sites of the RuO$_2$(110) lattice. The system is prepared in an initial state corresponding to (a) an empty lattice and (b) 1 ML oxygen coverage on cus sites and vacant bridge sites. The temperature is 450 K and the gas phase CO and O$_2$ pressures are both 1 bar.}
\label{fig:relaxation}
\end{figure*}

In Fig.\ \ref{fig:relaxation}(a) we show the temporal evolution of the system beginning from an initial state corresponding to an empty $(20\times20)$ lattice and at 1\,bar O$_2$ and CO pressure and a temperature of 450\,K. At these high pressures, it is clear that a substantial fraction of the surface will be covered with O or CO in the final steady state. Indeed, in the very first KMC steps (\ie on nanosecond timescales!), this coverage builds up quickly. The O coverage builds up roughly double as fast as the CO coverage, since every O$_2$ adsorption event leads to the appearance of two O atoms on the surface in contrast to only one CO in a CO adsorption event. Already after about 20-30\,ns one could be mistaken to assume that the system has reached the steady-state solution, since both the TOF and the surface coverages become constant. These coverages roughly reflect the impingement situation, with about 2/3 of all cus and br sites covered by O and about 1/3 of all cus and br sites covered by CO. However, this is not the true steady state! During prolonged KMC simulation (now on a timescale of seconds!) the coverages change again dramatically, until only about 1/3 of all br sites are covered by O (the rest by CO) and essentially all cus sites are covered by CO. In the course of these longer term changes, the TOF decreases by more than three orders of magnitude as compared to the premature apparent steady state. The reason for this two-timescale behavior is that the double as high impingement of O atoms onto the initially empty lattice fills a lot of br sites with oxygen. These O$^{\rm br}$ atoms are very strongly bound and quite unreactive. They will definitely not desorb readily and it requires CO oxidation reactions with comparably high barriers (small rate constants) to remove these O$^{\rm br}$ atoms once they have formed. The latter processes therefore take much longer than the initial filling, and since in this longer-term transformation CO also replaces almost all of the highly reactive O$^{\rm cus}$ species, the TOF also collapses.

As impressively demonstrated by this example, identifying what is the true steady state in a KMC simulation can be a major challenge. In fact, this holds even more since what one would really want is to have some form of automatized algorithm that would flag once a simulation has reached steady state. In practical applications, dozens, if not hundreds of different KMC simulations need to be conducted, for instance evaluating different gas phase conditions or determining the influence of changes in the catalyst surface composition (doping with certain metal atoms etc.). It becomes increasingly impractical and cumbersome to having to monitor central quantities of interest during an ongoing KMC simulation and then judge manually whether steady state has been reached. Unfortunately, fully automatic and fool-proof such steady-state detection (SSD) is not yet available for KMC simulations. In contrast, quite some knowledge is for instance available in the area of signal processing and process control \cite{Cao1995,Kelly2013}. Only very recently such algorithms have also found their way into the KMC community \cite{Nunez2017,Nellis2018}. Typically, they are applied to several properties of interest (reaction rates, coverages, total lattice energy etc.) in order to avoid a false-positive detection of the steady state. Even so, further testing and method development will probably be needed to ascertain whether they can always be applied in a foolproof manner.

In the absence of such sophisticated SSD algorithms, one pragmatic approach is often to have a lot of knowledge of the studied system and be really really cautious (certainly not a foolproof and elegant solution though). The other, complementary approach is to start simulations from mindfully chosen varying initial conditions and then monitor if the same steady state is reached. This implies that the system does not exhibit true multiple steady states. Such behavior is well known from the solution of differential equations, for instance in the context of MFA microkinetic models \cite{Ramachandran1981}. For KMC simulations in the context of surface catalysis such true multiple steady states that are not the result of a prematurely assumed convergence of the simulation have not been reported, and differentiating between the two cases would likely also be involved. In any case, it never hurts to run several KMC simulations starting from different initial conditions and then monitor where they converge to. Obviously, the closer the initialization is chosen to the final steady state, the faster the simulation will likely converge. 

Notwithstanding, one might also be trapped in preconceived configurations, when for instance initializing a simulation with a steady-state configuration determined in a preceding (similar) simulation. For the CO oxidation model we illustrate in Fig.\ \ref{fig:relaxation}(b) how starting from a completely different initial population does in fact lead to the same steady state. Here, the initial state corresponded to 1 ML oxygen coverage on the cus sites and vacant bridge sites. For nanosecond timescales we see again the appearance of a (different!) quasi-steady-state solution where the cus sites remain occupied by O, while the bridge sites are covered by about 0.35 ML CO and about 0.65 ML O.
However, similar as for the other initialization, on the timescale of seconds the system then transforms again and we arrive at the same steady-state solution as found there. Having understood the reason for this two-timescale behavior, a most suitable initialization would instead be to prepare all br sites with CO to prevent the initial massive buildup of O$^{\rm br}$. In this case, the true steady state is in fact reached extremely quickly (not shown). Yet, such detailed insight into the chemistry of the system is rarely present in the first place and the approach of starting from largely different initializations is often the best and hopefully reliable shot we have.

Once the steady-state solution has been reached, one can make use of the (hopefully present) ergodicity of the KMC simulation to calculate the desired quantities (surface composition, occurrence of various elementary steps, TOFs) as time averages instead of ensemble (trajectory) averages. The average reaction rate, $\langle r^{\beta} \rangle$, for the production of a given molecule $\beta$ can for instance be calculated as
\begin{equation}
\langle r^{\beta} \rangle = \frac{1}{t_{\rm KMC}} \sum_{n = 1}^{N_{\rm KMC}} \sum_{j} k_{ij}^{\beta} \Delta t_{\mathrm{escape},n} \quad,
\end{equation}
where $t_{\rm KMC}$ is the total KMC simulation time, the first sum runs over all KMC steps $n$ (up to a total of $N_{\rm KMC}$ steps), the second sum runs over all states $j$ that are accessible from the current state $i$, $k_{ij}^{\beta}$ is the rate constant for a process involving the production of the molecule $\beta$, and $\Delta t_{\mathrm{escape},n}$ is the escape time for KMC step $n$. The total simulation time should be chosen long enough to reduce the statistical error on the sampled quantities to a desired value. When calculating statistical errors over successive trajectory fractions, the fraction simulation time must also be chosen long enough that each trajectory fraction is statistically independent of the other fractions. The required time is known as the decorrelation time and it describes the time it takes before the current system configuration is uncorrelated from the initial system configuration in the trajectory fraction. In case the quantity of interest is the transient behavior of the system prepared in a given initial state (\eg to simulate the aforementioned temperature programmed reaction \cite{Jansen1995,Rieger2008}, cyclic voltammetry \cite{Rai2006} or titration \cite{Piccinin2014} experiments), several statistically independent trajectories must be calculated and averaged. The statistical independence can for example be achieved by using different random number seeds, as was done for the adatom diffusion model in Sec.\ \ref{sec:diff_example} above.

\section{Sensitivity analysis and uncertainty quantification}
\label{sec:sensitivity}

As already motivated in Sec.\ \ref{sec:diff_example} it can be of particular interest to know the extent to which the rate constants of the various processes in a model influence the model predictions and to thereby quantify the uncertainty of those predictions. This is generally known as sensitivity analysis and uncertainty quantification (UQ). Some of the main questions these methodologies seek to answer are the following: i) Error propagation: How do errors introduced at the level of theory used to calculate rate constants propagate to the model predictions? Which rate constants are most important to calculate with a high accuracy (maybe then recursively refine such calculations once the critical rate constants are known)? What conclusions can be reliably drawn despite the errors? ii) Design and optimization: What are the limitations to achieve an optimal performance of \eg a given catalyst or battery material? How should rate constants be varied (\eg by varying the material) to achieve such optimal performance?

One sensitivity measure introduced specifically for catalysis is Campbell's degree of rate control (DRC) $X_{{\rm RC},I}$ for reaction step $I$ \cite{Campbell1994}
\begin{equation}
X_{{\rm RC},I} = \frac{k_I}{\langle r^{\beta} \rangle} {\left( \frac{\partial \langle r^{\beta} \rangle}{\partial k_I} \right)}_{k_{J \neq I},K_I} \quad,
\end{equation}
where the average reaction rate $\langle r^{\beta} \rangle$ should be calculated for the product $\beta$ of interest and the derivative is evaluated holding constant the rate constants $k_J$ of all other reactions steps $J$ and the equilibrium constant $K_I$ for step $I$. A positive (negative) DRC signifies that the reaction rate will increase (decrease) when increasing $k_I$, whereas a value of zero signifies that the reaction rate is insensitive to variations in $k_I$. The DRCs follow a sum rule, which states that the sum of all DRCs is equal to one \cite{Meskine2009,Hoffmann2017}. A single non-zero DRC equal to one then signifies a single rate-limiting step in the reaction mechanism, while in general several steps can be rate-limiting at the same time. The fact that the equilibrium constant for step $I$ is held constant means that both the forward and reverse rate constants for step $I$ are varied simultaneously, which can also be viewed as a variation of the TS energy of step $I$ (for activated processes). Later, the DRC concept was extended to a thermodynamical version where instead the energies of reaction intermediates are varied \cite{Stegelmann2009}. Obviously, these sensitivity measures can easily be extended to other quantities of interest than reaction rates. From a practical point of view, the main challenge in evaluating the derivative entering the DRC expression is that we don't have an analytical expression for the reaction rate in KMC. Relying instead on a finite-difference sampling, very long simulation times are typically required to reduce the statistical error sufficiently \cite{Meskine2009}. A more efficient three-stage approach has recently been proposed \cite{Hoffmann2017}, which allows for the direct sampling of sensitivity measures from a single KMC trajectory.

DRC sensitivity measures are formulated as a linear response theory, meaning that the result is only valid locally in the input parameter space. However, kinetic models are most often highly non-linear and the DRC can change substantially over rate constant variations corresponding to a DFT error of \eg $\pm$0.2 eV on reaction barriers. Recently, a number of methods have therefore been developed for global sensitivity analysis and UQ, \ie to assess which conclusions about the model can be reliably drawn despite uncertainties in the input parameters \cite{Sutton2016,Dopking2017}. Ref.\ \cite{Sutton2016} furthermore addressed the fact that the errors in the input parameters for KMC models are often highly correlated. Such correlations can arise because the used DFT functional might generally over- or underestimate certain kinetic parameters. Corresponding DFT functional correlations have been exploited to assess the uncertainty of reaction rates in a MFA microkinetic model for ammonia synthesis \cite{Medford2014} carried out using the Bayesian error estimation functional with van der Waals correlation (BEEF-vdW) \cite{Wellendorff2012}. The latter functional provides not only a single value for a given kinetic parameter, but an ensemble of values generated by sampling known uncertainties in the exchange-correlation model parameters. 
Another source of correlations in kinetic parameters is the existence of correlations in the adsorption energies of chemically related intermediates and TSs, generally known as scaling relations \cite{Norskov2002,Michaelides2003,Abild2007} (see Sec.\ \ref{sec:ts_scaling}).
Both sources of correlation were considered in the sensitivity analysis and UQ carried out in Ref.\ \cite{Sutton2016} and applied to a kinetic model for ethanol steam reforming. The method allows to assess whether a proposed reaction mechanism can be considered to agree with experimental data within the known DFT errors and can reveal limitations such as the failure to take into account support effects in heterogeneous catalysis models.

To be quite honest, no such sophisticated sensitivity analysis is yet really on the agenda of the large majority of practitioners. As a very crude and simple advise in the context of this practical guide to surface Monte Carlo simulations, we would thus recommend to always at least vary some of the key rate constants in a KMC model by hand and see how this changes the simulation result. This is straightforward to do and it provides a crude (possibly incomplete, but still better than none) picture of what could be the critical input and what are the dependencies in the studied reaction network. In case this flags a critical sensitivity, one can and should escalate from there. Honestly, if the gist of the story one extracts from KMC simulations depends critically on highly specific numerical values of one or a few rate constants and one knows that there is a large uncertainty in these values, then who wants to stick their neck out?

\section{Timescale disparity problem}
\label{sec:timescale}

KMC achieves an enormous speedup over MD simulations by avoiding the explicit treatment of the vibrational degrees of freedom of the system, and instead considering only the rare events such as adsorption/desorption, diffusion or reaction steps (see Sec.\ \ref{sec:rare-events}). However, also those rare events that are treated in the KMC simulation can occur at largely different timescales. When this is the case, almost all of the CPU time is spent simulating the fast processes, while the (maybe truly important) dynamics arising from the slow processes is sampled insufficiently or not at all. This is especially problematic for KMC models of surface reactions on metals, where often very fast surface diffusion processes and slow surface reactions occur simultaneously in the reaction network.

A wide variety of methods has been developed to deal with this timescale disparity problem. The $\tau$-leap method is an approximate procedure in which the KMC simulation is accelerated by the firing of multiple processes at once \cite{Gillespie2001}. The underlying assumption (leap condition) of this method is, however, only fulfilled when surface populations are approximately constant during the coarse time increment $\tau$. Hence, the method does not apply to surface reactions on microscopic lattices, where site populations can change dramatically (\eg from zero species to one species) from one KMC step to the next. In practice, the method is only applied to coarse-grained lattices where the concentration of species within one larger coarse-grained cell is approximately constant in time. Other methods rely on a separation of the processes into "slow" and "fast" processes. They treat then only the slow process dynamics stochastically at the KMC level, while the fast process dynamics is treated deterministically or using a Langevin equation \cite{Haseltine2002,Salis2005}. A main drawback of this class of methods is that the process timescale separation has to be specified by the user in advance and remains fixed throughout the simulation.

\begin{figure}
\centering
\includegraphics*[width=0.5\textwidth]{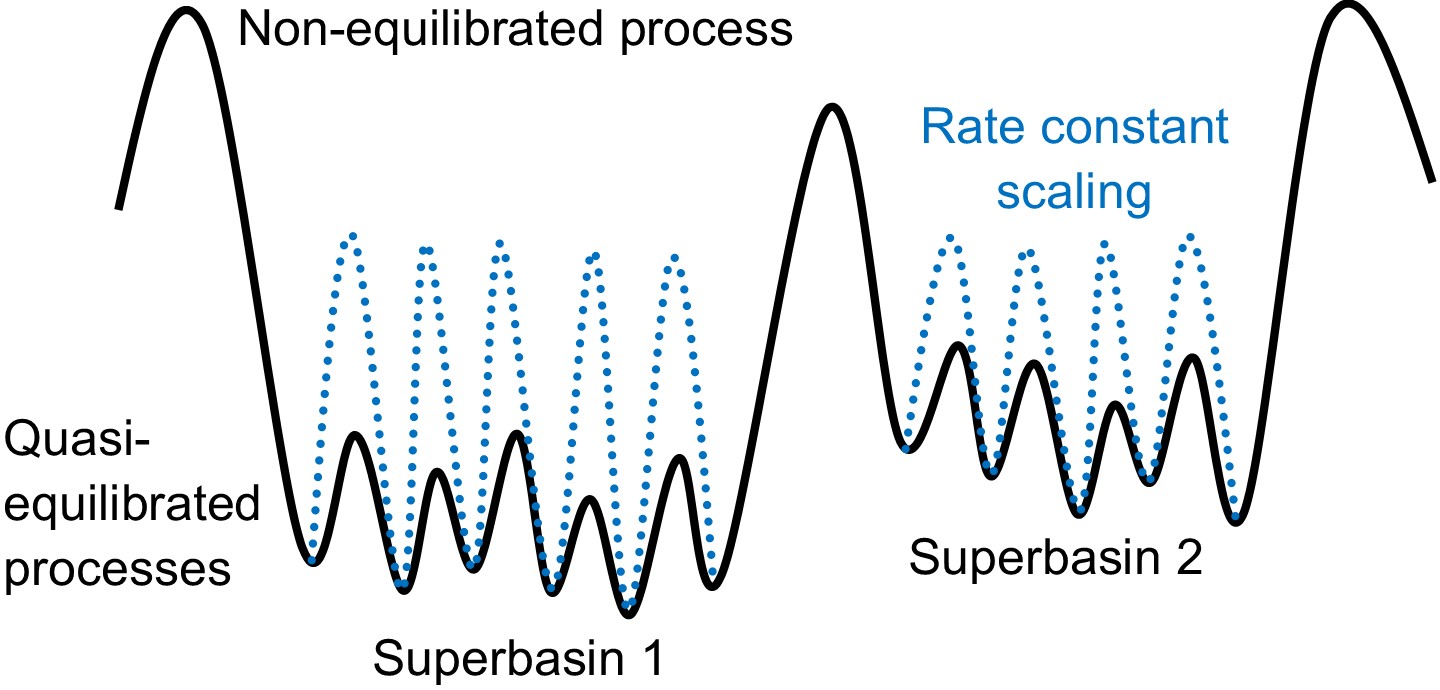}
\caption[]{Potential energy surface (PES) for a system with processes occurring at disparate timescales due to large differences 
in the barriers. A set of states connected by fast (low-barrier), quasi-equilibrated processes constitutes a superbasin. 
The system can escape from one superbasin to another through the execution of a slow (high barrier), non-equilibrated process. 
The KMC simulation is accelerated by scaling the rate constants (increasing the barriers) of fast, quasi-equilibrated processes. This decreases the time it takes for the system to leave the current superbasin and thereby enhances the sampling of neighboring regions of the PES (other superbasins).}
\label{fig:scaling}
\end{figure}

In practice, recent KMC works have therefore often relied on simple acceleration schemes that function by decreasing the rate constants of the fastest processes in the system in order to make the span in process timescales tractable, see Fig.\ \ref{fig:scaling}. For simple reaction networks such rate constant scaling, together with verification that the model output is not affected by the scaling, can be carried out manually \cite{Rogal2008,Piccinin2014,Lorenzi2016,Jorgensen2017}. Recently, algorithms have also appeared that take care of the scaling of fast processes automatically, without the user having to specify those processes in advance \cite{Chatterjee2010,Dybeck2017}. A main assumption of these methods is that fast processes become quasi-equilibrated after a limited number of executions, \ie it is assumed that continued simulation of these processes is not necessary to correctly describe the system dynamics. The accelerated superbasin KMC (AS-KMC) method from Ref.\ \cite{Chatterjee2010} defines a superbasin as a set of lattice configurations that the system can jump between through the execution of quasi-equilibrated processes only, see Fig.\ \ref{fig:scaling}. The execution of a non-equilibrated process then defines the entering of a new superbasin. The goal of the acceleration algorithm thereby becomes to encourage the system to leave the current superbasin at an earlier time through the scaling of fast, quasi-equilibrated processes. A drawback of this method for complex systems is that the total number of system configurations can be very large, which may cause the algorithm to become inefficient since the full sampling of even a single superbasin becomes exceedingly slow.

This problem was addressed in the method presented in Ref.\ \cite{Dybeck2017}, where rather than tracking both system configurations (superbasin states) and processes, only the executions of some user-defined reaction channels are tracked. A reaction channel could for example be the adsorption/desorption of some species at a given site type, independently of where on the lattice this reaction occurs, and the scaling of rate constants is then applied to the whole reaction channel. Scaling still only occurs for processes that have been executed a minimum number of times within the current superbasin and for which the number of forward executions is roughly equal to the number of reverse executions to within some tolerance. Once a non-equilibrated process was executed, the rate constants are unscaled again  to allow for sufficient sampling of the new superbasin -- and the process is started over again. The algorithm was shown to work well for a reaction model of Fischer-Tropsch synthesis over ruthenium nanoparticles \cite{Dybeck2017}. Very similar approaches to KMC acceleration, but excluding the unscaling step, have been followed in Refs.\ \cite{Nunez2017,Hoffmann2018}. In Ref.\ \cite{Nunez2017} the acceleration scheme was employed together with efficient sensitivity analysis beyond finite differences (similar to the method from Ref.\ \cite{Hoffmann2017} that was discussed in the preceding section) for improved sampling of sensitivity measures also in KMC models characterized by large disparities in the timescales.

In Ref.\ \cite{Andersen2017} the algorithm from Ref.\ \cite{Dybeck2017} was implemented in the {\tt kmos} code and applied to a trend study of CO methanation over stepped surfaces of the transition metal series using scaling-relation-based rate constant expressions. Also for this more complex reaction mechanism and the consideration of many different catalyst surfaces, the acceleration algorithm generally performed well. However, some challenging cases leading to a breakdown of the algorithm were also identified. A particularly problematic case for the algorithm was shown to be the occurrence of reactions between two low-coverage species, which are both produced in independent, quasi-equilibrated reaction steps. In this case the algorithm may scale the rate constants of the quasi-equilibrated steps too aggressively, leading to insufficient or no sampling of lattice configurations where the two low-coverage species are found at neighboring lattice sites, as required for their reaction. This problem is likely related to the fact that the algorithm from Ref.\ \cite{Dybeck2017} does not track system configurations and therefore cannot verify if all configurations within a superbasin have been sufficiently sampled.
A simple correction scheme that takes into account lattice configurations of the nearest neighbor sites in the definition of the reaction channel was proposed in Ref.\ \cite{Andersen2017}. This was shown to work well for the simple case of a reaction between two low-coverage species formed directly at neighboring sites. However, it does not apply if the low-coverage species are produced at distant lattice sites and rely on diffusion steps before the reaction step.

In some sense it is ironic that KMC is so particularly challenged by fast diffusion steps, considering that its effective competitor in form of MFA microkinetic models is challenged by slow diffusion steps. Even in the absence of lateral interactions, which would generally be used to argue in favor of the validity of the MFA, such slow diffusion processes can prevent the system to ever reach the well-mixed state assumed in the MFA. At metals, this was shown to happen at strongly binding step sites \cite{Andersen2017}, whereas at oxides this might arise simply from the higher diffusion barriers at these more open compound materials. For the KMC model of CO oxidation on RuO$_2$(110) discussed in Sec.\ \ref{sec:cat_example_1} a corresponding MFA breakdown could for instance be traced back to a relatively high barrier of approximately 1.6 eV for O diffusion \cite{Temel2007,Matera2011,Exner2015}. In this respect, this leaves essentially no system of interest in surface catalysis where one could generally expect mean-field kinetics to yield the right answers: At metals, the MFA is typically invalidated by strong lateral interactions at nearby sites (see next section), at compounds like oxides high diffusion barriers prevent a sufficient mixing of the adsorption layer. The problem is then that even though modern KMC implementations like those discussed in Sec.\ \ref{algo} have become dramatically more efficient, they are typically still more demanding than mean-field models. The timescale disparity problem adds significantly to this and even though most recent acceleration algorithms have become better, this problem is not yet fully solved. This still leaves many users to resort to MFA microkinetic modeling, even though it is likely not correct. Alternatively, algorithms that are intermediate between MFA and KMC in terms of accuracy and computational cost such as the quasichemical approximation have also been applied to catalytic reactions \cite{Hellman2007} and have in some cases been shown to reproduce KMC simulations at significantly reduced cost.

\section{Lateral interactions}
\label{sec:lat_int}

Lateral interactions are interactions between species adsorbed to a lattice. They can be either attractive or repulsive depending on the chemical nature of the involved species and the surface or bulk material defining the KMC lattice. Several recent studies have found that lateral interactions are essential to take into account for a correct description of the system dynamics \cite{Stamatakis2016,Jorgensen2017,Piccinin2017,Vignola2017,Hus2019}. As already discussed in Sec.\ \ref{sec:lattice}, lateral interactions are taken into account within the lattice approximation in KMC by assigning an individual hopping rate constant to each neighbor configuration (see Fig.\ \ref{fig:lat_int}). In case of repulsive interactions, the particle will be more likely to jump away from an initial configuration with occupied neighboring sites as compared to empty neighboring sites. This will cause the particles to spread out over the lattice to maximize the inter-particle distances. For attractive interactions, the situation is exactly opposite and the particles will show a tendency for island formation. Both cases can lead to the formation of ordered structures on the lattice. They may therefore lead to inaccuracies of mean-field treatments of the system kinetics, since the MFA assumes a random distribution of the particles without any correlations between sites \cite{Liu2016,Stamatakis2016,Pineda2017}.

In practice, lateral interactions can be accounted for in lattice KMC models through the general cluster expansion method \cite{Sanchez1984,Stampfl1999,Muller2003}. In this method, the lattice energy is expanded into a sum of discrete interactions (clusters) such as pairwise interactions, three-body interactions etc.\ through a lattice-gas Hamiltonian. For any adsorbate on the lattice, one would thus evaluate how many neighbors of which type sit at which kind of distances (nearest-neighbor, next nearest neighbor etc.). For each such neighbor the adsorption energy (or more generally rate constant) of an isolated particle on the lattice at this site type would be corrected for by a certain amount prescribed by the lattice-gas Hamiltonian. Summing up all these contributions defines the pairwise interactions. Then one looks up all possible motifs of two simultaneously present neighboring species, for which again there are energy (rate constant) corrections. This defines the three-body corrections to the pairwise interaction correction, and formally this goes on to higher and higher many-body interactions. While cluster expansions considering up to three- and four-body terms have been parametrized with DFT for simple systems \cite{Jansen2005,Zhang2007,Schmidt2012,Wu2012,Piccinin2014}, presently cluster expansions are typically truncated already after the first nearest neighbor pairwise interaction term in more complex systems with many species and site types in order to keep the computational cost tractable \cite{Stamatakis2011,Yang2013}. A particularly crude form of this are so-called site-blocking rules \cite{Hoffmann2014a,Lorenzi2016,Liu2016}, where strongly repulsive first neighbor interactions are simply modeled by suppressing any KMC processes that would lead to such immediately neighboring species. Furthermore, cluster expansions are typically used only for adsorbates in their stable and metastable adsorption sites, since taking into account also transition states, \ie changes in barrier heights due to lateral interactions, could make the DFT parametrization intractable. In the {\tt ZACROS} KMC code cluster expansion is built-in for adsorbates and the effect of lateral interactions on transition states is instead taken into account through an approximate Br{\o}nsted-Evans-Polanyi relation \cite{Nielsen2013}. A main drawback of the cluster expansion method is that the computational cost of both the DFT parametrization and the KMC simulation can quickly become intractable for complex systems. A benchmarking of the cluster expansion parametrized in Ref.\ \cite{Wu2012} showed that the computational cost of the KMC simulation increased with about 5 orders of magnitude when the number of clusters considered increased from 3 to 12 \cite{Nielsen2013}. Of course, this depends on the actual KMC implementation and algorithm used, and in particular codes like ZACROS are written to mitigate the additional costs when considering lateral interactions.

Despite the added cost, we emphasize again that lateral interactions often play an important role not only for surface diffusion, but also for crystal growth and heterogeneous catalysis. In the following we will thus provide two example {\tt kmos} models for the application areas crystal growth and heterogeneous catalysis that take into account lateral interactions. For more realistic growth and lateral interaction models we refer to the literature \cite{Fichthorn2002,Ruan2010,Shirazi2014}.\\

\subsection{{\tt kmos} models with lateral interactions}
\label{sec:lat_int_examples}

We will begin this section with a short description of two of the available {\tt kmos} backends and the way they each handle lateral interactions. The \emph{local smart} backend is the original backend and has been used as a basis and inspiration for the other backends. It was built with the implicit objective of offering the best run time possible at the expense of memory usage. For this reason,
a key element in this backend is a pre-calculated list of rate constants, \ie a \emph{rate catalog}. Together with an efficient local updating of the available processes after each KMC step, this makes it the most efficient backend when the number of different rate constants in the list is reasonable small. The \emph{local smart} backend implements the most simple treatment of lateral interactions, in which any new neighbor configuration defines a new KMC process with a rate constant associated with this particular configuration. There are several drawbacks of this approach for models with lateral interactions since these generally feature an exponentially growing number of processes with the number of interactions taken into account: (i) Several routines whose execution time scale with the total number of processes can become slow, (ii) the bookkeeping data structures, which scale in size with the total number of processes, can become too big for available memory, (iii) the size of the source code can become very large, making compilation very slow or even impossible due to memory requirements.

The \emph{on-the-fly} backend was constructed to alleviate the problems encountered in the \emph{local smart} backend and to enable KMC simulations of complex lateral interaction models. As the name implies, it calculates rate constants \emph{on-the-fly} instead of making use of a pre-calculated rate catalog. In the \emph{on-the-fly} backend the local environment is taken into account for each process through a \emph{bystander} list. Here a bystander is a neighboring site whose occupation influences the rate constant of the process according to the on-the-fly-calculated rate constant. The benefit of this approach is that the total number of processes in the model is constant with respect to the number of interactions taken into account. On the other hand, the drawback is that now the compute time of each KMC step scales linearly with the system size due to the need to scan through the lattice and sum up all rate constants explicitly in order to evaluate the total escape rate constant $k_{\rm tot}$ (Eq.\ \ref{eq:k_tot}).

\begin{figure*}
\centering
\includegraphics[width=1.0\textwidth]{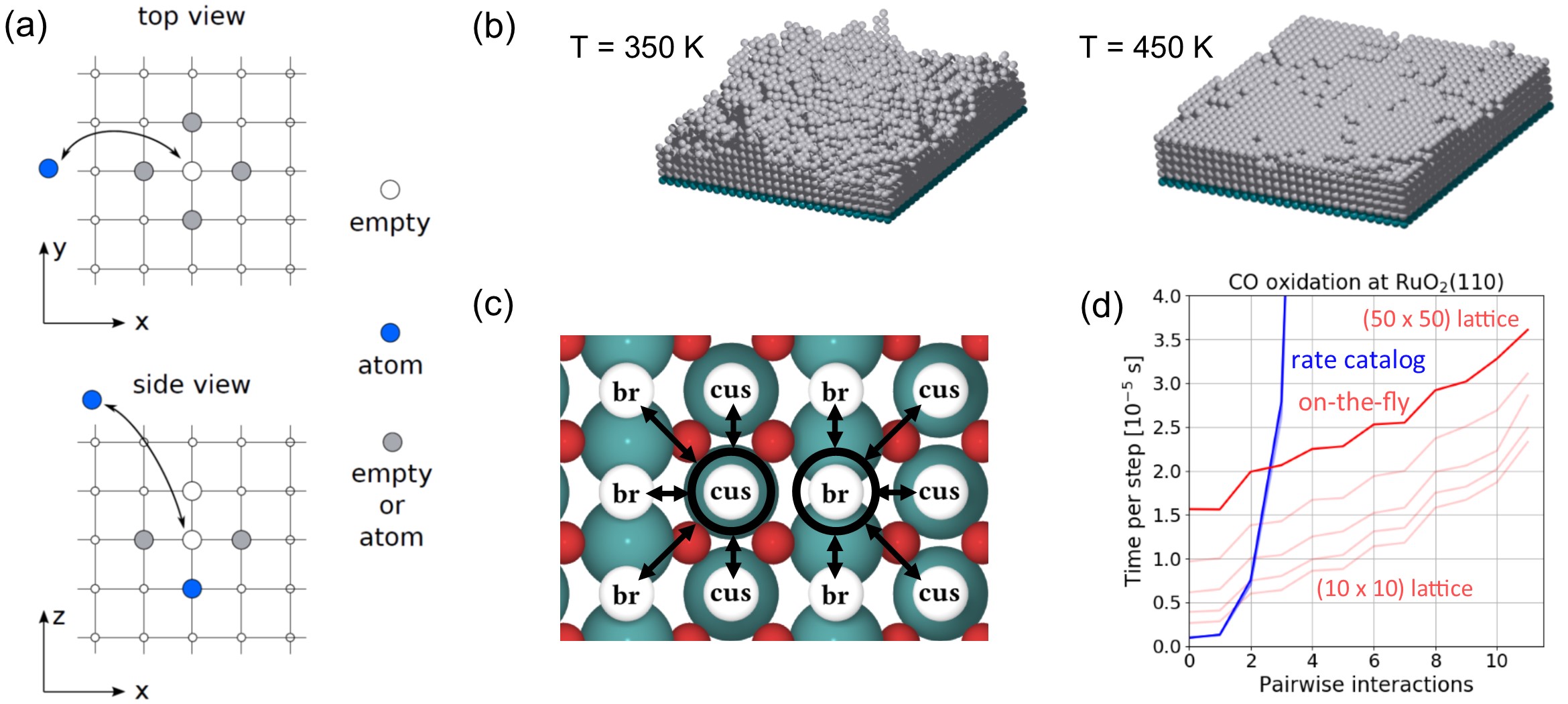}
\caption[]{(a) Illustration of the lateral interaction model for the desorption process of the crystal growth model. 
The site just below the desorbing species is always occupied and its interaction energy is therefore included in the low-coverage desorption barrier. The four neighboring sites at the same z height can be either empty or occupied and modify the desorption rate constant accordingly. (b) Snapshots of grown crystal structures at two different temperatures. (c) Illustration of pairwise interactions in the CO oxidation on RuO$_2$(110) model. (d) {\tt kmos} performance for the CO oxidation model as a function of the number of pairwise interactions considered for two different backends (rate catalog or on-the-fly calculation of rate constants). Using a rate catalog, the performance is independent of the lattice size. In the on-the-fly implementation the cost instead grows linearly with the lattice size (quadratic growth with the length $N$ of an $(N \times N)$ simulation cell) as illustrated for $N$ equal to 10, 20, 30, 40, 50 (different red lines).}
\label{fig:growth}
\end{figure*}

Now moving to examples, we will first present a {\tt kmos} model for crystal growth with lateral interactions that is simple enough that it can be handled efficiently in the \emph{local smart} backend. For this model we consider a three-dimensional quadratic lattice and a single species that is deposited onto a solid substrate with a constant adsorption rate constant $k_{\rm ads} = 3 \cdot 10^{-3}$ s$^{-1}$.
The low-coverage desorption barrier $\Delta E_0$ is set to 1\,eV. In addition, we consider attractive pairwise lateral interactions $\epsilon_{\rm int}$ = 0.5\,eV with the nearest neighbor species. Thereby, the rate constant for the desorption process $k_{\rm des}$ becomes
\begin{equation}
k_{\rm des} = \frac{k_{\rm B} T}{h} \exp\left(-{\frac{\Delta E_0 + n_{\rm NN} \cdot \epsilon_{\rm int}}{k_{\rm B} T}}\right) \quad,
\end{equation}
where $n_{\rm NN}$ is the number of nearest neighbor species in the lattice configuration. For the desorption process to be possible the species must be located in the surface layer, \ie the site just below it must be occupied and the site just above it must be empty.
Lateral interactions are then taken into account for the four neighboring sites at the same z height, see Fig.\ \ref{fig:growth}(a).
Since there are only two possible configurations for each neighbor site (empty or occupied) this leads to merely 16 inequivalent desorption process types and this model can therefore be efficiently treated in the \emph{local smart} backend.

In Fig.\ \ref{fig:growth}(b) we show snapshots of the grown crystal structure for temperatures of 350 K and 450 K. In both simulations the system is prepared in an initial state corresponding to one layer of fixed substrate species (blue atoms) onto which adsorption can take place. As expected, the grown structure becomes smoother at higher temperatures, where atoms deposited onto unfavorable adsorption sites with no attractive interactions to neighboring species will be more likely to desorb. The model could be made more realistic by including also diffusion processes and by considering a more detailed cluster expansion model for the lateral interactions.

As the next example we return to the catalysis model for CO oxidation on RuO$_2$(110) already discussed in Sec.\ \ref{sec:cat_example_1}. To explore how the {\tt kmos} performance of the \emph{local smart} and \emph{on-the-fly} backends are each affected by increasingly complex lateral interaction models, we step-wise add pairwise interactions to each process in the model. Possible pairwise interactions are illustrated with black arrows in Fig.\ \ref{fig:growth}(c) for a second-order process (\eg O$_2$ desorption) involving two neighboring sites (marked with black circles). The performance of each backend as a function of the number of interactions included is shown in Fig.\ \ref{fig:growth}(d). For the \emph{local smart} backend (rate catalog) the cost grows exponentially with the number of interactions and we are effectively limited to simple interaction models with a maximum of four pair-wise interactions. This happens because the number of processes in the rate catalog grows exponentially with the number of interactions, \ie the three possible occupations of each neighboring site (O, CO or empty) to the power of the number of neighboring sites (interactions) taken into account. In the \emph{on-the-fly} backend the cost grows only linear with the number of interactions and remains tractable at all considered system sizes, despite the quadratic scaling of the cost with the length $N$ of the $(N \times N)$ simulation cell used (linear scaling with system size). It is worth noting in this context that the \emph{on-the-fly} algorithm presented here specifically for {\tt kmos} is not new. For example, also the ZACROS code \cite{Nielsen2013} calculates rate constants \emph{on-the-fly}.

As has hopefully become evident from the above examples, lateral interactions are a double-edged sword for KMC models. On the one hand, the ability to systematically take into account effects of the local environment through systematic cluster expansion enables models of potentially very high accuracy and constitutes a great advantage over MFA models. On the other hand, the inclusion of lateral interactions counteracts the reduction in the number of needed first-principles rate constants that was obtained by making use of the lattice approximation (see Sec.\ \ref{sec:lattice}) and introduces additional complexity and cost to the KMC simulation. In practice, the choice of lateral interaction model therefore (unfortunately) often becomes a pragmatic one determined by what is computationally feasible, even if it is well-known that an appropriate interaction model can be crucial for the validity of the kinetic model.

\section{Summary and outlook}

The last decade has witnessed an impressive growth, not only in the number of studies employing predictive, first-principles KMC modeling, but also in the number of new codes that have become available. Especially for heterogeneous catalysis applications, the employed models are increasingly able to deal with complexity, both in the employed reaction mechanisms and in the structure of the catalytic material modeled. These advances have been powered by algorithmic developments for the determination of the input processes and rate constants, as well as for the actual algorithms that carry out the KMC simulation.

With many new KMC users entering this exciting field, the aim of this tutorial review has been to provide the necessary practical guidelines and examples for constructing and evaluating KMC models, and to highlight the pitfalls met along the way as well as current challenges and perspectives. We discussed in detail how to make use of the \emph{lattice approximation} in order to exploit a crystalline symmetry of the underlying surface lattice and thereby reduce the number of required first-principles rate constants. This also involves pitfalls and challenges, in particular when a dynamical restructuring of the surface takes place, or if well-defined adsorption sites of the surface species do not exist. We briefly discussed how \emph{off-lattice} (adaptive) KMC attempts to overcome these limitations, as well as the limitations associated with a pre-defined -- possibly incomplete -- list of possible processes, through the automatic identification of possible states and processes. However, despite this exciting perspective, these methods are still hampered by their high cost in practice.

Even for \emph{lattice} KMC models the number of required first-principles rate constants can be daunting. Prevalent TST approaches require the location of the TS and often also the determination of prefactors and zero-point vibrational energies, at least to the level of determining the harmonic frequencies of the system. While it is desirable to carry out these calculations by DFT or other electronic structure methods to achieve a predictive-quality model, the development of cheaper methods could dramatically increase the complexity that it is possible to tackle. To this end, we discussed commonly employed BEP and scaling relations as one prevalent example today. For the future, the development of increasingly accurate semi-empirical methods such as Density-Functional Tight-Binding (DFTB), reactive force fields or machine-learning-based force fields is an interesting perspective to which we ascribe a high potential. The actual determination of the TS is typically the computational bottleneck, and for this we discussed a number of methods and their respective strong and weak points. Furthermore, we discussed how the accuracy of the employed rate constants can be systematically improved in \emph{lattice} KMC models through the cluster expansion method in cases where the local environment significantly influences the rate constant.

For KMC models constructed using first-principles DFT rate constants, one still has to bear in mind that the expected error on barriers can easily be on the order of several hundreds of meV. This may lead to rate constants that are potentially wrong by orders of magnitude. A practical way to tackle this significant drawback is to estimate the sensitivity of the model predictions on the input rate constants and to quantify the uncertainties on those predictions. For this, we discussed various approaches ranging from a simple parameter variation to sophisticated models taking into account correlations between the input rate constants. An exciting perspective here is a recurrent refinement of the used rate constants, possibly also regarding the lateral interaction model employed, as information about the importance of the various input processes becomes increasingly available.

While KMC has come a long way, there are also a number of challenges to be met in the future. One example of this is the inevitable timescale disparity problem, which continues to challenge practical applications of KMC. We discussed a wide variety of methods that have appeared in the last decades, with a particular focus on recent acceleration algorithms that automatically identify the simulation bottlenecks, \ie the fast, quasi-equilibrated processes. While these approaches can work well in many cases, there are also examples where they break down. Likely, this is an area where further improvements are to be seen in the coming years.

\section{Conflict of Interest Statement}

The authors declare that the research was conducted in the absence of any commercial or financial relationships that could be construed as a potential conflict of interest.

\section{Author Contributions}

All authors listed have made a substantial, direct and intellectual contribution to the work, and approved it for publication.

\section{Funding}

This project has received funding from the European Union’s Horizon 2020 research and innovation programme under grant agreement 736299. Responsibility for the information and views set out in this article lies entirely with the authors. Chiara Panosetti would like to acknowledge funding from the German Research Foundation (DFG) through grant no DFG PA 2932/1-1. The publishing of this work was supported by DFG and the Technical University of Munich (TUM) in the framework of the Open Access Publishing Program.

\section{Acknowledgments}

The authors would like to acknowledge Juan Manuel Lorenzi and Michael Seibt for assistance in the development of the tutorial KMC models presented in this review article as well as for helping with creating some of the figures shown. 

\section{Supplemental Data}

All files required to set up and run the KMC models described in the text with the {\tt kmos} code as well as a text document with additional explanations are provided as supplemental data.

\bibliography{references}

\end{document}